\documentclass[aps,prx,twocolumn,citeautoscript,longbibliography]{revtex4-1}

\usepackage{amsmath,amssymb}
\usepackage{bm}
\usepackage{graphicx}
\usepackage{epstopdf}
\usepackage{latexsym}
\usepackage{subfigure}
\usepackage{color}
\usepackage{dsfont} 
\usepackage{wasysym} 
\usepackage{stmaryrd} 
\usepackage{hyperref} 

\definecolor{airforceblue}{rgb}{0.36, 0.54, 0.66}

\newcommand{\be}{\begin{equation}}
\newcommand{\ee}{\end{equation}}
\newcommand{\bea}{\begin{eqnarray}}
\newcommand{\eea}{\end{eqnarray}}

\begin{document}

\title{Probing Hidden Orders with Resonant Inelastic X-Ray Scattering}

\author{Lucile Savary}
\author{T. Senthil}
\affiliation{Department of Physics, Massachusetts Institute of
  Technology, 77 Massachusetts Ave., Cambridge, MA 02139}

\date{\today}
\begin{abstract}
We propose a general scheme for the derivation of the signals resonant
inelastic (and elastic) x-ray scattering (RIXS) gives access to. In
particular, we find that RIXS should allow to directly detect many
hidden orders, such as spin nematic, bond nematic, vector and scalar
spin chiralities. To do so, we choose to take the point of view of
effective operators, leaving microscopic details unspecified, but
still keeping experimentally-controllable parameters explicit, like
the incoming and outgoing polarizations of the x-rays. We ask not what
microscopic processes can lead to a specific outcome, but, rather,
what couplings are in principle possible. This approach allows to
systematically enumerate all possible origins of contributions to a
general RIXS signal. Although we mainly focus on magnetic insulators,
for which we give a number of examples, our analysis carries over to
systems with charge and other degrees of freedom, which we briefly address. We hope this work will help guide theorists and experimentalists alike in the design and interpretation of RIXS experiments.
\end{abstract}

\maketitle

\section{Introduction}
\label{sec:introduction}

Many systems have ground states with well-defined order parameters
which couple directly to conventional probes such as neutrons or
light. The accessible data usually comes in the form of ``structure factors,''
i.e.\ correlation functions of two ``elementary'' observables. Classic examples are
magnetically ordered states, e.g.\ ferromagnets and antiferromagnets
whose magnetic structure and fluctuations can be resolved by methods
like neutron scattering, muon spin resonance ($\mu$SR), nuclear
magnetic resonance (NMR) etc.. However, many of the
``exotic'' phases proposed by theorists do not fall into this
category. Some states exist, for example, which possess a well-defined local order
parameter, but still evade robust characterization using
``conventional'' probes. The order is then
commonly referred to as ``hidden.'' Typically, the
order parameters of such systems have quantum numbers which are
multiples of those which elementary particles give access
to when coupled linearly to the system. For example neutrons can excite $S=1$
magnons, but not $S=2$ excitations (owing to the dipolar coupling
between the neutron and electron's spins). Perhaps the simplest and best-known example of a hidden order
is that of spin quadrupolar (also called nematic) order \cite{penc2011}. In that case,
the expectation values of the spin projections, $\langle S_i^\mu\rangle$ (note the spins transform as ``dipoles'') are zero, but
those of ``quadrupolar'' operators, like $\langle S_i^\mu S_i^\nu\rangle$, are
not. Many other types of hidden orders have been proposed in the
literature
. Among
those are spin ``bond nematic'' \cite{chubukov1991,starykh2014}, where the order parameter contains spins
on neighboring sites, and spin vector and scalar chiralities, which involve
 antisymmetric products of spins. 
Hidden orders also arise in
conducting systems, with the famous 
example of nematic (in that case, ``nematic''
refers to rotation --discrete or continuous-- symmetry breaking in
real space)
order in the pnictide superconductors.

Here we show that Resonant Elastic and Inelastic X-Ray Scattering
(REXS and RIXS) can in
principle measure spin nematic, vector and scalar chirality, and many more
correlation functions (static and dynamical for REXS and RIXS, respectively). 
In general, we propose an enveloping scheme which allows to
systematically enumerate which correlation functions will contribute
to the RIXS signal in any given polarization geometry. REXS signals
are obtained from RIXS in the $\omega\rightarrow0$ limit. In
particular, in the case of static order, REXS signal should display
corresponding ``Bragg'' peaks.


``Resonant scattering'' refers to techniques where the energy of an
incoming probe is tuned to a ``resonance'' (a.k.a.\ ``edge''). In that
case, not only is the absorption (virtual or real) cross-section
dramatically enhanced, but the latter may also involve nontrivial
operators, allowing to probe correlation functions of complex order
parameters, i.e.\ typically those of hidden orders, which are
otherwise hardly accessible. This is clear upon thinking in terms of
perturbation theory in the probe-system coupling amplitude, and we
soon specialize to an x-ray probe. The
scattering amplitude up to second order is given by \cite{messiah1962,ament2011} 
\begin{equation}
  \label{eq:11}
  \mathcal{T}_{\mathtt{f}\mathtt{i}}=\langle\mathtt{f}|\hat{H}'|\mathtt{i}\rangle+\sum_n\frac{\langle\mathtt{f}|\hat{H}'|n\rangle
    \langle n|\hat{H}'|\mathtt{i}\rangle}{E_{\mathtt{i}}-E_n},
\end{equation}
where $|\mathtt{i},\mathtt{f}\rangle$ denote the initial and final
states of the \{system + electromagnetic (EM) field\}, $\hat{H}'$ is the
coupling Hamiltonian between matter and the EM field,
$\{|n\rangle\}$ forms a complete set of states (the ``important'' ones
will be discussed later) of the system, and
$E_\alpha$ is the energy of the state $|\alpha\rangle$. When there
exist states $|n\rangle$ which are close in energy to
$E_{\mathtt{i}}$, the system is said to be at resonance with the probe
and the second order amplitude in Eq.~\eqref{eq:11} largely dominates the
first. Moreover, within perturbation theory, the former
contains, among others, the following chain of (virtual) processes:
the absorption of a photon, the evolution of the resulting system,
followed by the emission of a photon. The RIXS signal is the cross-section relative to
the amplitude of such a process, when the incoming x-ray light is
tuned to a resonance which involves the excitation of a core electron
to a valence level, i.e.\ when $|n\rangle$ is a state of the pure
system (no photons) and contains a ``core hole''. Typical orders of
magnitude for such x-ray frequencies range between 0.01 and
10~keV \cite{dewey1969,bearden1967,ament2011}, i.e.\ correspond to
photon wavevectors of order 1-10$^{-3}$~\AA$^{-1}$. 

Detailed microscopic analyses of RIXS processes in a number of
systems have been described at length in the
literature \cite{ament2011}, some even predicting the observation of
correlation functions of complex order parameters \cite{ko2011,michaud2011}. Here we do
not belabor on them, but rather base the analysis entirely on the
observation that the initial (before the photon is absorbed) and final
states (after the photon is emitted) of the system both
belong to its low-energy manifold. 
Essentially, in that approach, the
only important feature of the microscopics is the reduction of (at least spatial)
symmetries to those of the core-hole site point
group. Such a symmetry-based strategy has a few major
advantages. An accurate description of all possible microscopic
processes is a very complex many-body problem, which is
moreover subjected to many uncertainties concerning the atomic
structure in a material. As a consequence, such approaches are
inherently material-specific. It is moreover very difficult to exhaust
all possible processes through microscopic reasoning. The symmetry procedure bypasses these issues.
This type of {\sl fully} effective approach was recently insightfully
pioneered in Ref.~\onlinecite{haverkort2010} in the context of
magnetic insulators, where the author gave
the form of on-site effective RIXS operators for up to two on-site spin operators. Here
we constructively rederive and generalize
Ref.~\onlinecite{haverkort2010}'s main result to all possible
symmetry-allowed couplings, including those which involve
multiple-site operators and degrees of freedom other than just spins. Moreover, the broader
context of the derivation presented here helps make more transparent
the correlations possibly probed in RIXS, on which we focus.

The remainder of this paper is organized as follows. We first review
the form of the light-matter interaction, the important symmetries to
be considered, and derive the form of the effective operators whose
correlations RIXS measures in
insulating magnetic systems, which are summarized in Table~\ref{tab:couplings}. We then turn to the study of three
important examples of hidden orders as may be realized in real
materials: spin nematic order, bond nematic order, vector and scalar
chiralities, and calculate the expected RIXS signals in these three
concrete cases. At the end of the paper we briefly
address systems with charge degrees of freedom.



\section{Effective operators}
\label{sec:effective-operators}

The leading
order Hamiltonian $\hat{H}'$ which couples light to matter and is
involved in the {\sl second}-order amplitude of the interaction
cross-section is given by, in the Coulomb gauge
${\boldsymbol{\nabla}}\cdot\mathbf{A}=0$ \cite{ament2011}\footnote{The
  ``diamagnetic'' term $\mathbf{A}^2$, of second order in
  $\mathbf{A}$, is involved in the
{\sl first}-order contribution to the scattering amplitude, which is
negligible close to resonance.}:
\begin{equation}
  \label{eq:8}
    \hat{H}'=\sum_\mathbf{r}\left[\hat{\psi}_\mathbf{r}^\dagger \,\frac{e\mathbf{p}}{m}\,\hat{\psi}_\mathbf{r}\cdot\mathbf{\hat{A}}_\mathbf{r}+\hat{\psi}_\mathbf{r}^\dagger\,\frac{e\hbar{\boldsymbol{\sigma}}}{2m}\,\hat{\psi}_\mathbf{r}\cdot\left({\boldsymbol{\nabla}}\times\mathbf{\hat{A}}_\mathbf{r}\right)\right],
\end{equation}
with the vector potential
\begin{equation}
  \label{eq:9}
  \mathbf{\hat{A}}_\mathbf{r}=\sum_\mathbf{k}\sqrt{\frac{\hbar}{2V\epsilon_0\omega_\mathbf{k}}}\sum_{{\boldsymbol{\varepsilon}}}\left({\boldsymbol{\varepsilon}}^*\hat{a}_{\mathbf{k},{\boldsymbol{\varepsilon}}}^\dagger
    e^{-i\mathbf{k}\cdot\mathbf{r}}+{\rm
    h.c.}\right).
\end{equation}
$\hat{H}'$ acts in the product space of the electrons
$\mathcal{H}_{e-}$ and photons $\mathcal{H}_{phot}$,
$\mathcal{H}=\mathcal{H}_{e-}\times\mathcal{H}_{phot}$,
$\hat{\psi}^\dagger$ and $\hat{\psi}$ are the electron creation and
annihilation second-quantized operator fields, $\hbar$ is Planck's
constant over $2\pi$, $e$ and $m$
are the electron charge and mass, respectively, $\hat{a}^\dagger$ and
$\hat{a}$ are the photon creation and annihilation operators,
${\boldsymbol{\varepsilon}}$ denotes the photon polarization, $V$ is the
volume in which the EM field is enclosed, $\epsilon_0$ is the
dielectric polarization of vacuum and $\omega_\mathbf{k}=\omega_{-\mathbf{k}}=c|\mathbf{k}|$ where $c$
is the speed of light. Here, for
concreteness, we make two approximations, often used in the
literature \cite{ament2011}: we consider {\sl (i)} that
$|\mathbf{k}\cdot\delta\mathbf{r}|\ll1$ at the relevant x-ray
wavelengths, where
$\mathbf{r}=\mathbf{R}+\delta\mathbf{r}$ where $\mathbf{R}$ denotes
the position of a lattice site, and so, at zeroth order,
$e^{i\mathbf{k}\cdot\delta\mathbf{r}}\approx1$ \footnote{For $\omega^{\rm x-ray}\sim10$~keV,
  $|\mathbf{k}|\sim1$~\AA$^{-1}$ and $|\mathbf{k}\cdot\delta\mathbf{r}|\approx0$ can seem hardly
  valid. In practice, however it has been shown to usually be a good
  approximation. Regardless, we discuss how to go beyond this approximation in Appendix~\ref{sec:higher-multipoles}.}, and {\sl (ii)} that in Eq.~\eqref{eq:8}
the magnetic term ($\propto{\boldsymbol{\sigma}}$) is subdominant compared to the ``electric'' one  ($\propto\mathbf{p}$). We
return to these approximations in
Appendix~\ref{sec:higher-multipoles}. Therefore, the second-order RIXS
amplitude for processes with a core hole at site $\mathbf{R}$
reduces to 
\begin{eqnarray}
  \label{eq:10}
\mathcal{T}_\mathbf{R}^{\mathtt{if}}&=& \sum_{\mathbf{q},\mathbf{q}',{\boldsymbol{\tilde{\varepsilon}}},{\boldsymbol{\tilde{\varepsilon}}}'}\left\langle
  \mathtt{f}\left|\left[{\boldsymbol{\tilde{\varepsilon}}}'\hat{a}_{\mathbf{q}'}e^{i\mathbf{q}'\cdot\mathbf{R}}+{\boldsymbol{\tilde{\varepsilon}}}'{}^*\hat{a}^\dagger_{\mathbf{q}'}e^{-i\mathbf{q}'\cdot\mathbf{R}}\right]\right.\right.\\
 &&\qquad\qquad\qquad\left.\left.\times\hat{\mathcal{O}}_\mathbf{R}\left[{\boldsymbol{\tilde{\varepsilon}}}\hat{a}_{\mathbf{q}}e^{i\mathbf{q}\cdot\mathbf{R}}+{\boldsymbol{\tilde{\varepsilon}}}^*\hat{a}^\dagger_\mathbf{q}e^{-i\mathbf{q}\cdot\mathbf{R}}\right]\right|\mathtt{i}\right\rangle\nonumber\\
 &=&\mathcal{A}_{\mathbf{k},\mathbf{k}'}\left\langle
  f\left|\varepsilon_\mu'{}^*\hat{\mathcal{O}}_\mathbf{R}^{\mu\nu}\varepsilon_\nu\right|i\right\rangle
 e^{i(\mathbf{k}-\mathbf{k}')\cdot\mathbf{R}},
\label{eq:10c}
\end{eqnarray}
where $\hat{\mathcal{O}}\sim
\frac{1}{\sqrt{\omega_\mathbf{q}\omega_{\mathbf{q}'}}}\mathbf{p}
 \,\hat{G}\,\mathbf{p}\,
 $, with
$\hat{G}=\sum_n\frac{|n_\mathbf{R}\rangle\langle n_\mathbf{R}|}{E_i+\hbar\omega_\mathbf{q}-E_n}$, where $|n\rangle$
are restricted to ``intermediate'' states with a core hole at site
$\mathbf{R}$ (i.e.\ close to resonance). The second
expression Eq.~\eqref{eq:10c} is obtained by requiring
$|\mathtt{i}\rangle=|i\rangle\otimes|\mathbf{k}{\boldsymbol{\varepsilon}}\rangle$
and
$|\mathtt{f}\rangle=|f\rangle\otimes|\mathbf{k}'{\boldsymbol{\varepsilon}}'\rangle$. Importantly,
$\hat{\mathcal{O}}$ acts purely in electronic space, and
moreover {\sl within the low-energy manifold}, provided the system
immediately ``returns'' to a low-energy state as the outgoing photon
is emitted, 
as is usually assumed. 
We therefore ask: what
effective operator acts purely in this
manifold which reproduces the matrix elements
$\mathcal{T}^{\mathtt{if}}_\mathbf{R}$? If we know the low-energy manifold and a basis which spans it, and if
the basis elements are physically meaningful, we shall immediately obtain
which correlation functions RIXS produces. We insist once again that, within this
approach, all ``intermediate processes,'' no matter how complicated,
are in a sense included, and need not be discussed.

As usual, most general arguments stem from symmetry considerations, which
we now address. The core hole is immobile, which imposes a strong symmetry constraint
on $\hat{\mathcal{O}}_\mathbf{R}$: it should be invariant in real
space under point (site $\mathbf{R}$) group symmetries. Another
constraint comes from the ``locality'' of the
effect of the core-hole in the ``intermediate propagation time''
$\tau=1/\Gamma\sim10^{-15}$~s \cite{ament2011},
which implies that only operators which act in close proximity to the
site of the core hole should be involved. While this statement may
appear somewhat loose, a quick order-of-magnitude analysis shows that,
{\em even in a metal}, electrons will not travel over more than very
few lattice spacings over the time $\tau$ \footnote{Indeed,
$\tau\sim10^{-15}$~s corresponds to an energy of order $4$~eV, which
is typically that of a metal's bandwidth $W$. Taking estimate of an
electron's velocity as $v=aW$ with
$a$ the lattice spacing, we find that the travelled distance
during time $\tau$ of order a lattice spacing.}. Finally, since transition amplitudes are scalars, by keeping the
polarization dependence explicit, we impose constraints
on the combination of operators which multiply the polarization components. This is what we address now and is summarized in Table~\ref{tab:couplings}.

For
concreteness and ease of presentation of the derivation we now focus
on magnetic insulators, though we note that the same ideas carry over to
systems with charge (and other) degrees of freedom, to which we return at the end
of the paper, in Sec.~\ref{sec:other-degr-freed}. 
Indeed, because of the ``locality'' of the effective scattering
operators, insulating systems are more readily tackled. Local (in the sense of acting only on degrees of freedom
living in a small neighborhood in real space)
operators in insulating systems yield a very natural description of
the system, and the low-energy manifold, being
finite (generally a well-defined $J$ multiplet, possibly split by
crystal fields) and sharply defined (usually a gap separates multiplets), can be spanned by effective
``spin'' operators (finite vector spaces of identical dimensions are
isomorphic). Therefore only a spin operator basis compatible with the
combinations of polarizations remains to be found. 


In the absence of both spin-orbit coupling {\sl at low energies} (core
levels always experience very strong spin-orbit
coupling \cite{ament2011}) and of a magnetic field, the system
should be rotationally symmetric in spin space. Moreover, in
  principle, in the Hamiltonian, under spatial symmetries, the spins
  are left invariant. However, here, in the RIXS structure factor, the
situation is more subtle. Spin excitations (and hence spin operators) may only arise in the
structure factor thanks to
spin-orbit coupling at the core. Therefore, in principle the structure
factor itself should display signs spin-orbit coupling \cite{haverkort2010b,haverkort2010}, with the effective spin
operators transforming under {\sl lattice} symmetries. Even upon
neglecting transition operators which break rotational symmetry if spin orbit
coupling is weak at the valence level, the effective spins still
transform under {\sl real space} symmetry
operations.\footnote{More rigorously, one should derive the transition
  operators in terms of spin-orbit coupled effective spins, and {\sl
    then} possibly neglect those which are not rotationally symmetric.} Then, the
polarizations and (effective) spins (the latter make up the
operators $\hat{\mathcal{O}}^{\mu\nu}$, as mentioned above) transform as
usual vectors and pseudo-vectors,
respectively, under spatial transformations, and according to
${\boldsymbol{\varepsilon}}\rightarrow-{\boldsymbol{\varepsilon}}^*$ and
$\mathbf{S}\rightarrow-\mathbf{S}$ under time reversal (see
Appendix~\ref{sec:transformation-rules}). In other words, under the
full spherical symmetry group, using the notations from Ref.~\onlinecite{snoke}, ${\boldsymbol{\varepsilon}}$ and $\mathbf{S}$
transform under $D_1^-$ and $D_1^+$, respectively (under
$SO(3)$ operations, both the polarization and spin vectors transform under the $L=1$
representation). Since $D_1^\pm\times D_1^\pm=D_0^++D_1^++D_2^+$ ($1\times
1=0+1+2$ for $SO(3)$), {\sl any} combination of spin operators which transform
under the same representations can in principle be involved in the
RIXS signal. Depending on the number of neighboring operators one
chooses to include (and on the
value of $S(S+1)$), possibilities differ. The situation for up to
three spin operators (on the same or nearby sites, from ``locality'') is
summarized in Table~\ref{tab:couplings} (see in particular the caption), and details of the derivation
are given in Appendix~\ref{sec:derivation-table-i}.
\begin{table*}[htb]
\centering
  \begin{tabular}{c|c||c|c|c}
 representation & polarizations & one spin & two spins & three spins\\
\hline
0 & ${\boldsymbol{\varepsilon}}'{}^*\cdot{\boldsymbol{\varepsilon}}$ &
& $\mathbf{S}_i\cdot\mathbf{S}_j$&
$\left(\mathbf{S}_i\times\mathbf{S}_j\right)\cdot\mathbf{S}_k$\\
1 & ${\boldsymbol{\varepsilon}}'{}^*\times{\boldsymbol{\varepsilon}}$
& $\mathbf{S}_i$ & $\mathbf{S}_i\times\mathbf{S}_j$ &
$\left(\mathbf{S}_i\cdot\mathbf{S}_j\right)\mathbf{S}_k$,\quad
$\left(\mathbf{S}_i\times\mathbf{S}_j\right)\times\mathbf{S}_k$,\quad $
\left\llbracket\mathbf{S}_i,\mathbf{S}_j\right\rrbracket\cdot\mathbf{S}_k$\\
2 & $\llbracket{\boldsymbol{\varepsilon}}'{}^*,{\boldsymbol{\varepsilon}}\rrbracket$
& & $\llbracket\mathbf{S}_i,\mathbf{S}_j\rrbracket$ &  $ \left\llbracket\mathbf{S}_i\times\mathbf{S}_j,\mathbf{S}_k\right\rrbracket$,\quad $\left\llbracket\mathbf{S}_i,\mathbf{S}_j\right\rrbracket\times\mathbf{S}_k$
  \end{tabular}
\caption{{\sl Generic} form of the operators in magnetic systems which couple to combinations of the
  polarizations in the absence of spin-orbit coupling. The double
  brackets represent the traceless symmetric products, $\llbracket \mathbf{u},\mathbf{v}\rrbracket_{\mu\nu}=
  \frac{1}{2}\left(u^\mu v^\nu+u^\nu
    v^\mu\right)-\frac{1}{3}(\mathbf{u}\cdot\mathbf{v})\delta_{\mu\nu}$, and the dot and vector
  products between a matrix (obtained through
  $\llbracket.,.\rrbracket$) and a vector are defined such that: $(\llbracket\mathbf{u},\mathbf{v}\rrbracket\cdot\mathbf{w})_\mu=\sum_\nu\llbracket\mathbf{u},\mathbf{v}\rrbracket_{\mu\nu}w_\nu$,
  $(\llbracket\mathbf{u},\mathbf{v}\rrbracket\times\mathbf{w})_{\mu\rho}=\sum_{\nu,\lambda}\epsilon_{\nu\lambda\rho}\llbracket\mathbf{u},\mathbf{v}\rrbracket_{\mu\nu}w_\lambda$ 
 (see Appendix~\ref{sec:derivation-table-i}). Moreover, the product
 $\llbracket\mathbf{u},\mathbf{v}\rrbracket\llbracket
 \mathbf{w},\mathbf{t}\rrbracket$ also denoted
 $\llbracket\mathbf{u},\mathbf{v}\rrbracket\cdot\llbracket
 \mathbf{w},\mathbf{t}\rrbracket$ is defined to be the fully symmetric
 product with all indices contracted: $\sum_{\mu\nu}\llbracket\mathbf{u},\mathbf{v}\rrbracket_{\mu\nu}\llbracket
 \mathbf{w},\mathbf{t}\rrbracket_{\mu\nu}$. Each row corresponds to a
given irreducible representation of a combination of the incoming and
outgoing polarization, given in the second column. Each entry on the
right of the double bar gives the combinations of spins which
transform as does the combination of polarizations on the same
line. The columns simply indicate the number of spin operators
involved in the effective operator. In
  principle, RIXS may measure correlations functions of the operators
  given in this table. This table is also ``valid'' for matrices which
connect local ``band'' indices with the same symmetries in conducting
systems. See Sec.~\ref{sec:conclusions}.}
\label{tab:couplings}
\end{table*}

{\sl On-site terms.---} Upon considering on-site terms
only ($i=j=k$), where one need not take into account any further lattice
symmetries, and {\sl up to two} spin operators, we recover the expression from
Ref.~\onlinecite{haverkort2010}:\footnote{Note that
  Ref.~\onlinecite{haverkort2010} additionally provides a relation
  between some of the coefficients $\alpha_\beta$ and absorption spectroscopy coefficients.}
\begin{equation}
  \label{eq:15}
  T_i=\alpha_0({\boldsymbol{\varepsilon}}'{}^*\cdot{\boldsymbol{\varepsilon}})+\alpha_1({\boldsymbol{\varepsilon}}'{}^*\times{\boldsymbol{\varepsilon}})\cdot\mathbf{S}_i+\alpha_2\llbracket{\boldsymbol{\varepsilon}}'{}^*,{\boldsymbol{\varepsilon}}\rrbracket \llbracket\mathbf{S}_i,\mathbf{S}_i\rrbracket,
\end{equation}
where $T_i=\varepsilon'^*_\mu\mathcal{O}_i^{\mu\nu}\varepsilon_\nu$,
and where $\llbracket \mathbf{S}_i,\mathbf{S}_j\rrbracket$ is the traceless
symmetric second rank tensor constructed from  $\mathbf{S}_i$ and
$\mathbf{S}_j$, i.e.\ given by: $\llbracket \mathbf{S}_i,\mathbf{S}_j\rrbracket_{\mu\nu}=
  \frac{1}{2}\left(S_i^\mu S_j^\nu+S_i^\nu
    S_j^\mu\right)-\frac{1}{3}(\mathbf{S}_i\cdot\mathbf{S}_j)\delta_{\mu\nu}$,
and analogously for
$\llbracket{\boldsymbol{\varepsilon}}'{}^*,{\boldsymbol{\varepsilon}}\rrbracket$. The symmetric
product
$\llbracket{\boldsymbol{\varepsilon}}'{}^*,{\boldsymbol{\varepsilon}}\rrbracket
\llbracket\mathbf{S}_i,\mathbf{S}_i\rrbracket=\sum_{\mu,\nu}\llbracket{\boldsymbol{\varepsilon}}'{}^*,{\boldsymbol{\varepsilon}}\rrbracket_{\mu\nu}
\llbracket\mathbf{S}_i,\mathbf{S}_i\rrbracket_{\mu\nu}$ has all
indices contracted. The $\alpha_n$ are material-specific coefficients \cite{haverkort2010}. The generalization to
discrete symmetries is {\sl formally} straightforward (though usually gruesome in practice) and discussed in
detail in Appendix~\ref{sec:lower-symmetry}. 

{\sl Off-site terms.---} The above considerations take care of the symmetry aspects relative
to spin space. To fulfill the constraints associated with the lattice,
which enters through $\mathbf{S}_\mathbf{r}\rightarrow
[\det R]\,R\cdot\mathbf{S}_{R\cdot\mathbf{r}}$ where $R$ is a spatial operation
(see Appendix~\ref{sec:transformation-rules}), the expressions must
be appropriately symmetrized. For example, take a 1d chain of $S=1/2$, and consider a maximum of two spin terms. Then, if lattice
sites are centers of inversion, the transition
operator will be (still assuming spherical symmetry in spin space):
\begin{eqnarray}
  \label{eq:16}
  T_i&=&\alpha_0 ({\boldsymbol{\varepsilon}}'{}^*\cdot{\boldsymbol{\varepsilon}})\mathbf{S}_i\cdot(\mathbf{S}_{i-1}+\mathbf{S}_{i+1})\nonumber\\
&&+({\boldsymbol{\varepsilon}}'{}^*\times{\boldsymbol{\varepsilon}})\cdot\left(\alpha_{1,1}\mathbf{S}_i+\alpha_{1,2}\mathbf{S}_i\times(\mathbf{S}_{i-1}+\mathbf{S}_{i+1})\right)\nonumber\\
&&+\alpha_2 \llbracket{\boldsymbol{\varepsilon}}'{}^*,{\boldsymbol{\varepsilon}}\rrbracket\llbracket\mathbf{S}_i,\mathbf{S}_{i-1}+\mathbf{S}_{i+1}\rrbracket,
\end{eqnarray}
where the $\alpha_n$ and $\alpha_{n,m}$ are material-specific coefficients which
multiply terms which belong to the same irreducible representation ($n$) (or
copy ($m$) thereof if an irreducible representation appears multiple times).

From Table~\ref{tab:couplings}, one may directly read out the
quantities whose correlation functions will contribute to the RIXS
signal, as well as which polarization geometry will reveal them while
switching off (most of) the other contributions (e.g.\
${\boldsymbol{\varepsilon}}'{}^*\parallel{\boldsymbol{\varepsilon}}$
will ``switch off'' the
${\boldsymbol{\varepsilon}}'{}^*\times{\boldsymbol{\varepsilon}}$ ``channel''). Indeed the
differential cross-section is given by \cite{messiah1962}
\begin{eqnarray}
  \label{eq:14}
  &&\frac{\delta^2 \sigma}{\delta\Omega\delta
    E}\nonumber\\
&&\propto\sum_{f}\left|\sum_{\mathbf{R},\mathbf{q}}\langle
    f|T_{\mathbf{q}}|i\rangle
    e^{i(\mathbf{q}+\mathbf{k}-\mathbf{k}')\cdot\mathbf{R}}\right|^2\delta(E_f+\omega_{\mathbf{k}'}-E_i-\omega_{\mathbf{k}}) \nonumber\\
&&\propto\sum_\mathbf{q}\langle
i|T_{-\mathbf{q}}T_\mathbf{q}|i\rangle\delta(\mathbf{q}+\mathbf{k}-\mathbf{k}')\delta(\Delta
E-\omega_\mathbf{q}),
\end{eqnarray}
where $\delta\Omega$ and $\delta E$ denote elementary solid angle
(related to the momentum transfer
$\widehat{\mathbf{k}-\mathbf{k}'}$) and energy, respectively, and
where $\Delta E$ is the measured energy transfer.

Before moving on to the discussion of specific examples, we make a
couple of important remarks. {\sl (i)} It is important to note that, for
effective spin-$1/2$ systems, only off-site terms can
contribute to, for example, the
$\llbracket{\boldsymbol{\varepsilon}}'{}^*,{\boldsymbol{\varepsilon}}\rrbracket$
channel. Indeed, there exist only four (counting the identity)
linearly independent $S=1/2$ operators. Therefore, while off-site
contributions are expected to be weaker (they may only arise from so
called ``indirect'' processes \cite{ament2011}), in an effective
$S=1/2$ system, a ``multi-site'' signal in the
$\llbracket{\boldsymbol{\varepsilon}}'{}^*,{\boldsymbol{\varepsilon}}\rrbracket$
channel will not ``compete'' with signal from possibly-larger onsite
couplings, offering hope to unambiguously detect such correlations. {\sl (ii)} We caution
that, of course, this symmetry-based approach does not any give information on the
absolute or relative strengths of the signals in the different channels. Moreover, ``selection
rules'' relative to the chosen ``edge'' need to be additionally taken
into account. {\sl (iii)} An additional word of caution is in order:
as far as we understand, the measurement of the {\sl outgoing}
polarization is not currently possible on instruments being used at
this point, although the new state-of-the art facility currently under
construction (which will also provide much higher resolution in
energy, currently at around $100$~meV) will be able to. 

\section{Spin nematic in the bilinear-biquadratic $S=1$ model on the triangular lattice}
\label{sec:spin-nemat-bilin}

The $S=1$ bilinear biquadratic model with Hamiltonian 
\begin{equation}
  \label{eq:9}
  H=\sum_{\langle i,j\rangle}\left(J_1\mathbf{S}_i\cdot\mathbf{S}_j+J_2(\mathbf{S}_i\cdot\mathbf{S}_j)^2\right),
\end{equation}
on the triangular lattice has been quite intensively studied,
especially so in recent years after it was suggested that it could be
relevant to the insulating material NiGa$_2$S$_4$, where Ni$^{2+}$ is
magnetic, with
$S=1$ \cite{nakatsuji2005,tsunetsugu2006,lauchli2006,bhattacharjee2006,stoudenmire2009,kaul2012}.
This material is made of stacked triangular planes of Ni$^{2+}$ ions,
and displays no long-range spin ordering but low-temperature specific
heat which grows with temperature as $T^2$ \cite{nakatsuji2005}. The
latter facts motivated the minimal description of NiGa$_2$S$_4$ by the
model Eq.~\eqref{eq:9}, which, for $J_1>0$, features two quadrupolar phases, one ``ferroquadrupolar'' and one
``antiferroquadrupolar.'' These phases are characterized by a
vanishing expectation value for the spins, $\langle S_i^\mu\rangle=0$,
but an on-site ``quadrupolar'' (a.k.a.\ ``spin nematic'') order
parameter: $\langle\{S^\mu_i,S^\nu_i\}-2\delta_{\mu\nu}\rangle\neq0$
(a diagonal part is subtracted to obtain a traceless operator). 
Since here we look not to accurately make
predictions for the actual material NiGa$_2$S$_4$, but rather to
demonstrate that RIXS will provide unambiguous signatures of
quadrupolar order, we now restrict our attention to the minimal
bilinear-biquadratic model Eq.~\eqref{eq:9}, despite the fact that the latter
will clearly not account for all the experimental features (not discussed here) of
NiGa$_2$S$_4$ \cite{stoudenmire2009}. 

\begin{figure}[htb]
  \centering
  \includegraphics[width=3.3in]{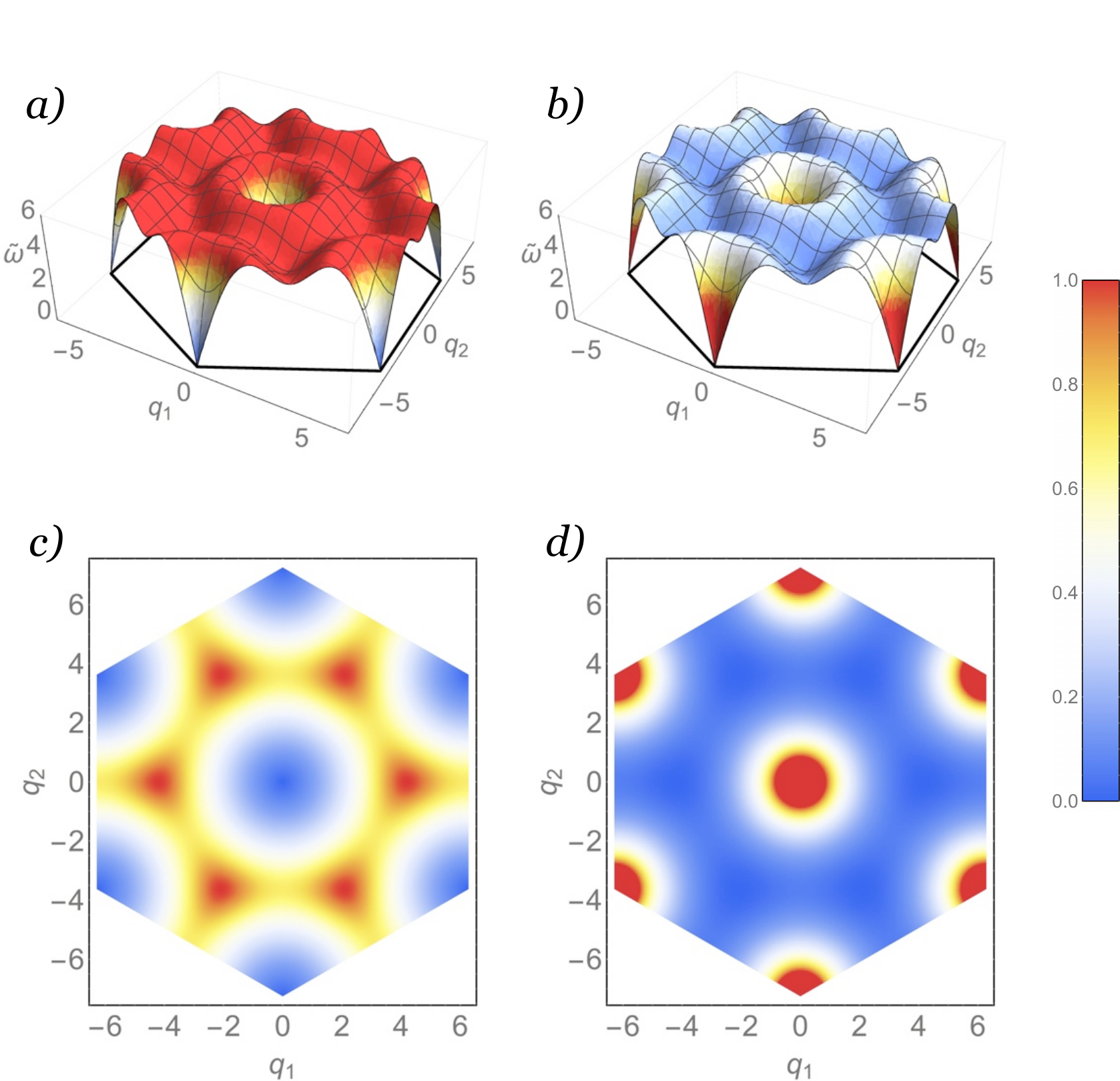}
  \caption{Color plots of the a) spin-spin correlation function (as
    probed by e.g.\ inelastic neutron
    scattering or RIXS with
    $\boldsymbol{\varepsilon}'{}^*\perp\boldsymbol{\varepsilon}$ and
    $\alpha_2$ small enough), and b) signal probed by RIXS for
    ${\boldsymbol{\varepsilon}}=(i,1,0)/\sqrt{2}$ and
    ${\boldsymbol{\varepsilon}}'=(1,i,0)/\sqrt{2}$, both for the model
    of Eq.~\eqref{eq:9} with $J_2/J_1=-\tan(7\pi/16)$ (ferroquadrupolar
    phase \cite{lauchli2006}) on the triangular
    lattice. c) and d) Equal time (integrated over frequency) versions
    of the signals shown on plots a) and b), respectively. Note that the intensities are {\sl independently} normalized. On figure b), the intensity is seen to diverge at the
    Goldstone mode, in sharp contrast with the vanishing of the
    spin-spin correlation function at the same point in figure
    a). In plots a) and b), $\tilde{\omega}=\omega/\sqrt{J_1^2+J_2^2}$.}
  \label{fig:1}
\end{figure}

The wavefunctions
of nematic states are simple single-site product wavefunctions. For
spin-1 systems, product wavefunctions can generally be expressed as
$|\psi\rangle=\prod_i\mathbf{d}_i\cdot|\mathbf{r}_i\rangle$, where
we have defined
$|\mathbf{r}_i\rangle=(|x_i\rangle,|y_i\rangle,|z_i\rangle)$, where
$\mathbf{d}_i\in\mathbb{C}^3$ and $|\mathbf{d}_i|=1$. 
The states $|\mu_i\rangle$ are time-reversal invariant states defined such that
$S_i^\mu|\mu_i\rangle=0$, i.e., in terms of the usual eigenstates of the
$S_i^z$ operator, $|x\rangle=\frac{i}{\sqrt{2}}(|1\rangle-|\overline{1}\rangle)$,
$|y\rangle=\frac{1}{\sqrt{2}}(|1\rangle+|\overline{1}\rangle)$ and
$|z\rangle=-i|0\rangle$ \cite{smerald2013}. In the case of a ``pure'' quadrupolar phase,
for this basis choice (with time-reversal invariant states),
$\mathbf{d}_i\in\mathbb{R}^3$ \cite{smerald2013}, which one can check indeed leads to
$\langle\mathbf{S}_i\rangle=\mathbf{0}$. The vector $\mathbf{d}_i$ at
each site is called the ``director,'' and corresponds to the direction along which
the spins do {\sl not} fluctuate. In nematic states, the direction along which the
director points may vary at each site, like in the
``antiferroquadrupolar'' phase of the above model, where the {\sl directors}
form a three-sublattice 120$^\circ$ configuration. In the ferroquadrupolar phase, the directors on each site point in the same
direction, which can be arbitrarily (since the Hamiltonian is
isotropic in spin space) taken to be the $z$ direction. In that case,
the unit cell is not enlarged. 
In ordered (or field-polarized) ferromagnets and
antiferromagnets, the low-energy elementary excitations of the system
are spin flips/waves, i.e.\ $S^z=\pm1$ local excitations. In nematic
states, where it is the directors which are ordered, spin waves
translate to ``flavor waves'' where there are now two pairs of
conjugate ``transverse'' bosons. 
Flavor wave spectra and dipolar and
quadrupolar correlations for the model Eq.~\eqref{eq:9} on the triangular
lattice have been calculated in several references \cite{lauchli2006,penc2011,pires2014,voll2015}. Our derivation is
provided in Appendix~\ref{sec:spin-nematic}, and here we give the
full RIXS structure factor for the model, assuming on-site spin
operators only (expected to provide the largest contributions to the signal), and spherical 
symmetries (a derivation is provided in Appendix~\ref{sec:lower-symmetry}), and
provide a few plots in Figure~\ref{fig:1} for various polarization geometries and
assumptions on relative absorption coefficients (about which symmetry
analysis gives no further information).
\begin{eqnarray}
  \label{eq:10}
&&\mathcal{I}_{\omega,\mathbf{q}}^{\rm RIXS}\propto
\sqrt{\frac{A_{\mathbf{q}}^2}{A_{\mathbf{q}}^2-B_{\mathbf{q}}^2}}\left[\left(\kappa_{xy}^{(2)}{}^2+\kappa_{yz}^{(2)}{}^2\right)\left(1-\frac{B_{\mathbf{q}}}{A_{\mathbf{q}}}\right)\right.\\
&&\left.\qquad\qquad\qquad+\left(\kappa_z^{(1)}{}^2+\kappa_x^{(1)}{}^2\right)\left(1+\frac{B_{\mathbf{q}}}{A_{\mathbf{q}}}\right)\right]\delta\left(\omega-\omega_{\mathbf{q}}\right)\nonumber,
\end{eqnarray}
where $A_\mathbf{q}=\frac{1}{2}(J_1\gamma_\mathbf{q}-6J_2)$,
$B_\mathbf{q}=\frac{\gamma_\mathbf{q}}{2}(J_2-J_1)$,
$\omega_\mathbf{q}=2\sqrt{A_\mathbf{q}^2-B_\mathbf{q}^2}$ with
$\gamma_\mathbf{q}=2\left(\cos
  q_1+\cos(\frac{1}{2}[q_1+\sqrt{3}q_2])+\cos(\frac{1}{2}[q_1-\sqrt{3}q_2])\right)$
and
$\kappa^{(1)}_{\mu}=\alpha_1\epsilon_{\mu\lambda\rho}\varepsilon^\lambda\varepsilon'{}^*{}^\rho$ ($\epsilon$ is the
second rank fully antisymmetric tensor),
$\kappa^{(2)}_{\mu\nu}=\alpha_2(-2/3\delta_{\mu\nu}({\boldsymbol{\varepsilon}}'{}^*\cdot{\boldsymbol{\varepsilon}})+\varepsilon'{}^\mu{}^*\varepsilon{}^\nu+\varepsilon{}^\mu{}^*\varepsilon'{}^\nu{})$
(note that $\alpha_1$ and
$\alpha_2$ depend, in particular, on the details
of the atomic and crystal structures \cite{haverkort2010}), see Appendix~\ref{sec:spin-nematic}. Quadrupolar correlations are therefore {\sl
directly} seen. Clearly, one recovers the proper scaling of the
amplitudes for the Goldstone mode (the system
spontaneously breaks spin-rotation symmetry in the ferroquadrupolar
phase) at $\mathbf{q}=\mathbf{0}$ at low energy,
$\omega_\mathbf{q}\sim|\mathbf{q}|$ and $\mathcal{I}^{\rm RIXS,ferro}\sim1/\omega_\mathbf{q}$. Figure~\ref{fig:1} illustrates the associated
smoking
gun evidence for quadrupolar order provided by RIXS.

\section{Bond nematic and vector chirality in nearest and next-nearest
  neighbor $S=1/2$ Heisenberg chains in a field}
\label{sec:bond-nematic-vector}

The $S=1/2$ ferromagnetic nearest-neighbor and antiferromagnetic next-nearest-neighbor Heisenberg
model on a chain 
\begin{equation}
  \label{eq:12}
  H=\sum_i\left(-J_1\mathbf{S}_i\cdot\mathbf{S}_{i+1}+J_2\mathbf{S}_i\cdot\mathbf{S}_{i+2}-hS_i^z\right)
\end{equation}
with $J_{1,2}>0$ is thought to be a minimal model for LiCuVO$_4$, a
distorted ``inverse
spinel'' (with chemical formula ABB'O$_4$) material such that the system can be seen in a first
approximation as a set of parallel edge-shared CuO$_2$
chains separated by Li and V atoms \cite{enderle2005,hagiwara2011,mourigal2012}. Cu$^{2+}$ are magnetic ions with spin $1/2$. As will be important later, we
note that the point group symmetry at each Cu site contains inversion symmetry.  
This material displays a complex phase diagram, which is now believed
to show, from low to high field: incommensurate helical order,
spin density wave order along the chains, and, possibly, right below
the saturation field, a spin nematic state. 
Why the $J_1-J_2$ Heisenberg model of Eq.~\eqref{eq:12} seems like a reasonable starting point to
describe this material may be articulated as follows: {\sl (i)}
there is experimental evidence for chain structure physics (see
above), {\sl (ii)} Cu usually displays weak spin orbit
coupling, suppressing any strong anisotropy in spin space, and {\sl (iii)}
further-neighbor interactions in such compounds are usually
sizable, owing to the configurations of the exchange paths. In fact,
$J_1$ and $J_2$ were estimated to be $19$~K and $44$~K, respectively,
using neutron diffraction and susceptibility data on single
crystals \cite{enderle2005,nawa2014}. Moreover, in some parameter regime, a number of the phases
numerical simulations obtain for the model are reminiscent of those
experimentally observed in LiCuVO$_4$, as we now discuss.

For $J_2/J_1>1/4$, in a non-zero but weak enough field, the minimal {\sl model} has been shown to exhibit a nonzero vector spin chirality
$\mathbf{\hat{z}}\cdot{\boldsymbol{\chi}}_{i,{i+1}}=\mathbf{\hat{z}}\cdot(\mathbf{S}_i\times\mathbf{S}_{i+1})$ and
$\mathbf{\hat{z}}\cdot{\boldsymbol{\chi}}_{i,{i+2}}=\mathbf{\hat{z}}\cdot(\mathbf{S}_i\times\mathbf{S}_{i+2})$ (a non-zero $z$-component of the chirality does
not break any continuous symmetry of the model in a field applied
along the $z$-axis and is therefore allowed),
reminiscent of the helical order in LiCuVO$_4$. More
precisely, DMRG and exact diagonalization have probed signs of
long-range chirality correlations \cite{kolezhuzk2005,hikihara2008,sudan2009}, and the bosonization of the field
theory---which unveils a Luttinger liquid phase--- predicts
$\langle{\boldsymbol{\chi}}_{i,{i+1}}\cdot\mathbf{\hat{z}}\rangle\neq0$ and
$\langle{\boldsymbol{\chi}}_{i,{i+2}}\cdot\mathbf{\hat{z}}\rangle\neq0$
\cite{hikihara2008,mcculloch2008,sudan2009}.
The higher-field phase of the model numerically shows evidence of (bond)
quadrupolar correlations \cite{chubukov1991,hikihara2008,sudan2009,starykh2014}. 

Again, here we claim not to provide a detailed description of the
material, but we propose that RIXS might be able to probe vector chirality as well
as bond-nematic order in this system.

In order to compute the RIXS signal, we proceed like in
Ref.~\onlinecite{hikihara2008} closely follow their derivation, and start
from the limit $J_1\ll J_2$ of two decoupled chains (each with lattice
spacing $2a_0$). Each one may then
be independently bosonized. We use the conventional notations for the
boson fields, $\theta_{1,2}$ and $\phi_{1,2}$, where the indices are
chain labels, and
$[\phi_\nu(x',\tau'),\partial_x\theta_\mu(x,\tau)]=i\delta_{\mu\nu}\delta(x-x')\delta(\tau-\tau')$,
for $\mu,\nu=1,2$. The
spin operators are given by \cite{kolezhuzk2005,hikihara2008,mcculloch2008,starykh2014}
\begin{eqnarray}
  \label{eq:22}
  S^+_{\mu}(x)&=&e^{i\sqrt{\pi}\theta_\mu(x)}\\
&&\times\left((-1)^jb+b'\sin(2\pi
    Mj+\sqrt{4\pi}\phi_\mu(x))\right)\nonumber\\
S^z_{\mu}(x)&=&M+\frac{1}{\sqrt{\pi}}\partial_x\phi_\mu(x)\\
&&\qquad-(-1)^ja\sin(2\pi
Mj+\sqrt{4\pi}\phi_\mu(x)),\nonumber
\end{eqnarray}
where $x$ is the coordinate of a site, while $j\in\mathbb{Z}$ labels a
``unit cell'' of two sites $\{1,2\}$ (sites can be labelled by $l=2j+\mu$), $M$ is the total magnetization (due to the field),
and $a,b,b'$ are non-universal constants. Note that here the subscript
$\mu$ in $S_\mu^\alpha$ is unrelated to the subscript $i$ in Eq.~\eqref{eq:12}.
As mentioned above, when
$J_1=0$, the two chains are decoupled and each one obtains two
free-boson theories, with the action $\mathcal{S}_{\rm eff}^\mu=\int
dx\int d\tau
[\frac{v}{2}\left(K(\partial_x\theta_\mu)^2+\frac{1}{K}(\partial_x\phi_\mu)^2\right)+i\partial_x\theta_\mu\partial_\tau\phi_\mu]$, where
$K$ and $v$ are the Luttinger liquid parameter and spin velocity
of
the antiferromagnetic Heisenberg ($J_2$) spin chain in a field. $J_1\neq0$
introduces couplings between the chains. Then it is useful to
define $\gamma_\pm=(\gamma_1\pm\gamma_2)/\sqrt{2}$ for
$\gamma=\theta,\phi$. The coupling actions are $\mathcal{S}_1=g_1\int
dx\sin\left(\sqrt{8\pi}\phi_-+\pi M\right)$ and $\mathcal{S}_2=g_2\int
dx(\partial_x\theta_+)\sin\sqrt{2\pi}\theta_-$, where $0\leq
M\leq1/2$ is the magnetization per site, with parameters $g_1=-J_1a^2\sin\pi M$ and
$g_2=-J_1\sqrt{\pi}b^2/\sqrt{2}$, which will lead to ``bond nematic'' and
``vector chiral'' phases. 
This model displays scale invariance, and renormalization
group (RG) ideas apply. Then, within this
approach, if $g_2/g_1$ flows to zero (resp.\ infinity) under the RG flow where
high-frequency modes are integrated out,
the
system goes into the nematic, where $\phi_-$ gets pinned to a value
which minimizes the integrand of $\mathcal{S}_1$, (resp.\ vector
chiral, where it is the integrand of $\mathcal{S}_2$
which acquires a finite expectation value)
phase \cite{hikihara2008}. Details are given in
Appendix~\ref{sec:vect-chir-bond}.  

\begin{figure}[htb]
  \centering
  \includegraphics[width=3.3in]{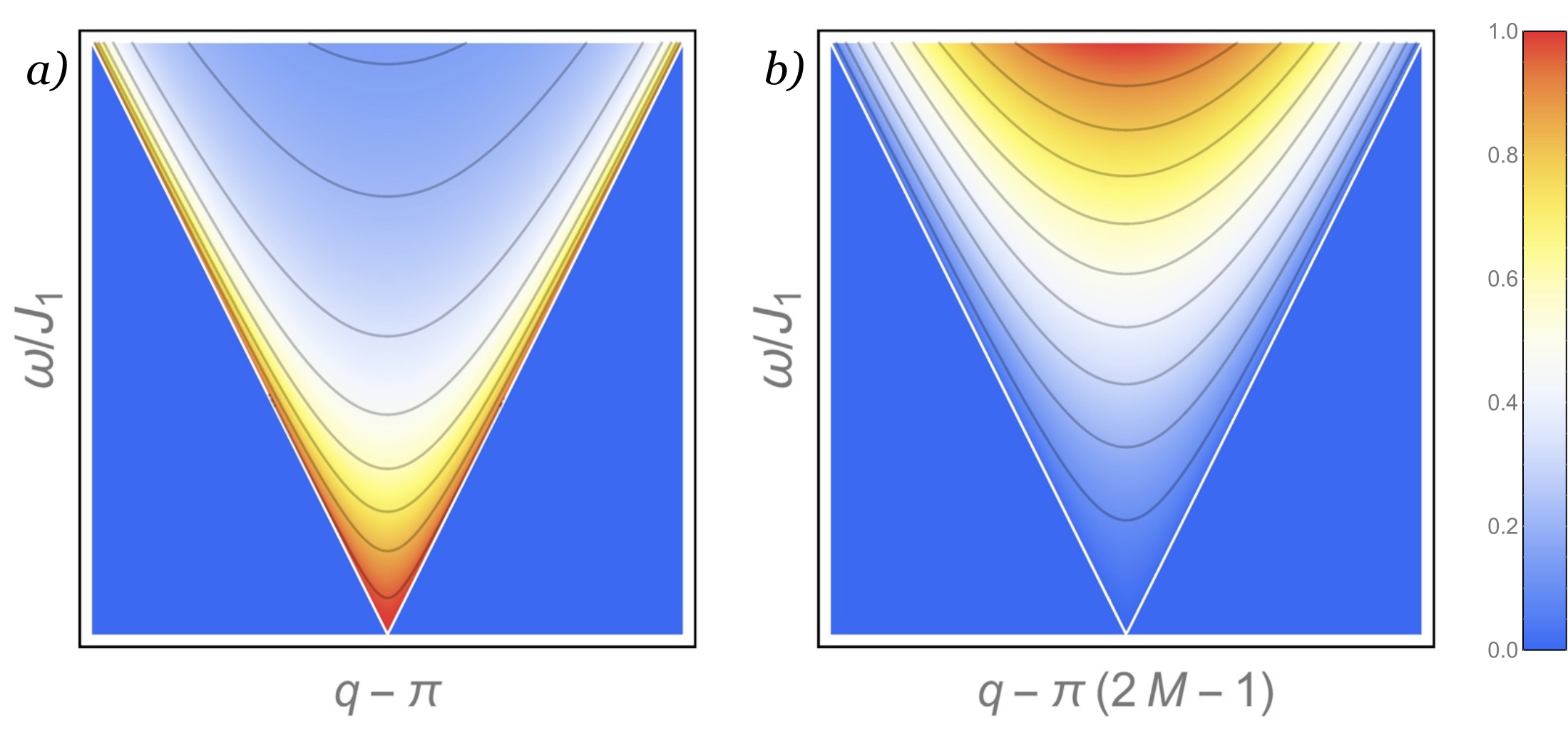}
  \caption{Color plots of the dominant contributions to the a)
    $\llbracket{\boldsymbol{\varepsilon}}'{}^*,{\boldsymbol{\varepsilon}}\rrbracket^\perp$ channel (fourth line of Eq.~\eqref{eq:24}) in the
    nematic phase, around $q=\pi$, as given in Eq.~\eqref{eq:27} b)
    $({\boldsymbol{\varepsilon}}'{}^*\times{\boldsymbol{\varepsilon}})^z$
    channel (third line of Eq.~\eqref{eq:24}) in the vector-chiral
    phase, around $q=\pi(2M-1)$, as given
    in Eq.~\eqref{eq:28}.}
  \label{fig:2}
\end{figure}

Because each site on the chain has only two neighbors, we expect that
the contributions to the RIXS signal from three-spin interactions
should be extremely weak. So, from Table~\ref{tab:couplings}, assuming
a weak enough effect of spin-orbit coupling at the low-energy level, the RIXS transition
operator is given by Eq.~\eqref{eq:16} in zero field, and by
\begin{eqnarray}
  \label{eq:24}
  T_i&=&\alpha_{0,\perp}
  ({\boldsymbol{\varepsilon}}_\perp'{}^*\cdot{\boldsymbol{\varepsilon}}_\perp)\mathbf{S}_i^\perp\cdot(\mathbf{S}_{i-1}^\perp+\mathbf{S}_{i+1}^\perp)\\
&&+\alpha_{0,z} ({\varepsilon}_z'{}^*{\varepsilon}_z)S_i^z(S_{i-1}^z+{S}_{i+1}^z)\nonumber\\
&&+({\boldsymbol{\varepsilon}}'{}^*\times{\boldsymbol{\varepsilon}})^z\left(\alpha_{1,1,z}\mathbf{S}_i+\alpha_{1,2,z}\mathbf{S}_i\times(\mathbf{S}_{i-1}+\mathbf{S}_{i+1})\right)^z\nonumber\\
&&+\alpha_{2,\perp}
\llbracket{\boldsymbol{\varepsilon}}'{}^*,{\boldsymbol{\varepsilon}}\rrbracket^\perp\llbracket\mathbf{S}_i,\mathbf{S}_{i-1}+\mathbf{S}_{i+1}\rrbracket^\perp,\nonumber
\end{eqnarray}
for $h\neq0$, i.e.\ when the full $SU(2)$ symmetry in spin space is broken down to
$U(1)$ (see Appendix~\ref{sec:lower-symmetry}). In Eq.~\eqref{eq:24},
we used the definitions
$\mathbf{u}=\mathbf{u}_\perp+u^z\mathbf{\hat{z}}$ and
$\llbracket\mathbf{u},\mathbf{v}\rrbracket^\perp_{\mu\nu}=\frac{1}{2}(u^\mu
v^\nu+v^\nu u^\mu)-\frac{1}{2}(\mathbf{u}_\perp\cdot\mathbf{v}_\perp)\delta_{\mu\nu}$
with $\mu,\nu=x,y$ only. The $\alpha_{n,\mu}$ and $\alpha_{n,m,\mu}$ are coefficients. 
Finally, we find the following {\sl low-energy} (long distance and
time) {\sl leading} contributions (see
Appendix~\ref{sec:vect-chir-bond}) to the RIXS structure factor:
\begin{eqnarray}
  \label{eq:27}
  \mathcal{I}_{\omega,q}^{\rm
    nematic}&\propto&\sum_{\epsilon=\pm1}\frac{\Theta(\omega^2-v_+^2(q-\epsilon\pi)^2)}{\sqrt{\omega^2-v_+^2(q-\epsilon\pi)^2}^{2-1/K_+}},
\end{eqnarray}
for, e.g., ${\boldsymbol{\varepsilon}}\times
{\boldsymbol{\varepsilon}}'{}^*=\mathbf{0}$ and
${\boldsymbol{\varepsilon}}\perp\mathbf{\hat{z}}$ in the nematic phase, and, around $q=\pm2\pi
   M\pm\pi$:
\begin{eqnarray}
  \label{eq:28}
  \mathcal{I}_{\omega,q}^{\rm
  chiral}&\propto&
   \sum_{\epsilon,\epsilon'=\pm1}\Theta(\omega^2-v_+^2(q-2\pi\epsilon
   M-\epsilon'\pi)^2)\\
&&\qquad\qquad\times\sqrt{\omega^2-v_+^2(q-2\pi\epsilon
   M-\epsilon'\pi)^2}^{4K_+-2}\nonumber
\end{eqnarray}
in cross polarizations, with $({\boldsymbol{\varepsilon}}\times
{\boldsymbol{\varepsilon}}'{}^*)\parallel\mathbf{\hat{z}}$ in the
vector chiral phase. 
In the expressions above, $K_+=K(1+J_1\frac{K}{\pi v})$ and
$v_+=v(1-J_1\frac{K}{\pi v})$ [note that $K(M=0)=1/2$ and $K(M=1/2)=1$]. Figure~\ref{fig:2} displays some examples.

\section{Other degrees of freedom: electrons, phonons and orbitals}
\label{sec:other-degr-freed}

The derivation of effective RIXS operators presented above in the
context of magnetic models readily extends to systems where other
degrees of freedom are important. Indeed, the symmetry
arguments we employed are general enough that they carry over to any type of
problem. 

Modifications arise at the level of the identification and choice of
basis for the space of operators which act on the local low-energy
manifold. In magnetic insulators, as discussed earlier, the natural
degrees of freedom are on-site, and a Hamiltonian is always associated
with the specification of what the local degrees of freedom, namely
effective ``spins,'' are. More microscopically, one can see an
effective spin degree of freedom ``emerge'' from the multiplet
structure of a single-ion Hamiltonian at each site. Now, similarly, if orbital degrees
of freedom are to be treated explicitly in an insulating system in
RIXS, one may simply introduce a set of (effective) operators
$\mathbf{L}$, $L^\mu L^\nu$ etc., which transform as pseudo-vectors
under {\sl real space} operations, and obtain a table similar to
Table~\ref{tab:couplings}, where now each row should be associated
with an irreducible representation of the appropriate point group.

Now, systems with charge degrees of freedom, or phonons, are usually
approached from a more field-theoretic perspective, where one has lost
sight of a microscopic model, and operators are labeled by some
momentum index  (among
others). That being said, given a material, one may
always, much like for the insulating magnet case, think about how
many electrons, and which single-ion orbital (or spin-orbital), a
given 
ion will ``contribute/provide'' to the valence band of
the whole solid. Provided one can determine this, it is reasonable to
think of these spin-orbital states and number of electrons as the
building blocks for the local low-energy manifold relevant to RIXS,
and the basis of operators can be made of those which reshuffle the
electrons in the (single-electron) spin-orbital states (even if the
electrons interact, such a non-eigenstate basis can be chosen
nevertheless). As an example, consider an atom A contributes $n$ on-site states to the valence band(s)
of the system, with creation and annihilation operators
$\psi_{\mathbf{r}\alpha}^\dagger,\psi_{\mathbf{r}\alpha}$. One may
build on-site operators $\psi_{\mathbf{r}\alpha}^\dagger
M_{\alpha\beta}\psi_{\mathbf{r}\beta}$,
$\psi_{\mathbf{r}\alpha}^\dagger \psi_{\mathbf{r}\beta}^\dagger
M_{\alpha\beta\gamma\delta}\psi_{\mathbf{r}\gamma}\psi_{\mathbf{r}\delta}$
etc., where $\alpha,\beta,\gamma,\delta=1,..,n$ (may be orbital
labels, for example), as well as some
involving neighbors, 
$\psi_{\mathbf{r}\alpha}^\dagger
M_{\alpha\beta}\psi_{\mathbf{r}'\beta}$, $\psi_{\mathbf{r}\alpha}^\dagger \psi_{\mathbf{r}'\beta}^\dagger
M_{\alpha\beta\gamma\delta}\psi_{\mathbf{r}'\gamma}\psi_{\mathbf{r}\delta}$,
etc.. Despite the more delocalized nature of the electrons in an
itinerant system, a quick order of magnitude estimate shows that, even
in a typical metal, only close-neighbor operators are involved in the
RIXS transition operators (see Sec.~\ref{sec:effective-operators} and
foonote therein). An additional
constraint in RIXS is charge conservation, since no electrons are
kicked out of the sample. Then, much like in the case of magnetic
insulators, we may split the tensors $M$ into irreducible
representations and obtain the coupling terms to the corresponding
combinations of polarizations. In a single band model, for example,
the only on-site operators are the density
$\psi^\dagger_\mathbf{r}\psi_\mathbf{r}$ and spin
$\psi^\dagger_{\mathbf{r}}{\boldsymbol{\sigma}}\psi_\mathbf{r}$ \cite{marra2013} (and
powers thereof, though the latter should be expected to contribute
sub-dominantly). 

Like in any endeavor to compare experiment with theory, in any other technique, the most-delicate step in the
{\sl calculation} of a structure factor in a given ground state will be to understand how the $\psi_{\mathbf{r}\alpha}$ operators from the
basis act on this ground state and are related to quasiparticle
operators, if any. This is particularly true in the case of metals
(but also of course in that of, e.g., quantum spin liquids),
where, even in the case of a Fermi liquid, where the notion of
quasiparticles is meaningful, the quasiparticle operators
$\Psi^\dagger$ are, in the crudest approximation, related to the
electron operators through the square-root of the quasiparticle weight
$0<Z\leq1$:
$\Psi^\dagger\sim\sqrt{Z}\psi^\dagger$. Therefore, a factor of at
least $Z^2$
will be involved in the contribution of a quasiparticle-related
excitation to the RIXS cross-section. Because $Z$ can be very small,
like in a highly correlated metal, it is important to keep track of
those factors to estimate the (esp.\ relative) amplitude of a signal
of a given origin. For example, upon taking the {\sl a minima} point of view
of a single-band Fermi liquid \cite{benjamin2014} for the
overdoped cuprates, one should keep in mind that factors of $Z$ are
likely to greatly suppress the quasiparticle contribution to the RIXS
signal. This should be crucial in deciphering the origin of the
features seen in RIXS spectra of those materials \cite{letacon2011,letacon2013,dean2013,wakimoto2015}.

The case of phonons is quite similar. At the symmetry level, phonons bear no spin degree of freedom, but are
associated to lattice degrees of freedom and their symmetries. There may be
several phonon/displacement modes at each site, so that one can
introduce several phonon creation operators $c^\dagger_{\mathbf{r},a}$. The symmetries to
be considered should be purely spatial, and related to point group
symmetries at site $\mathbf{r}$. 
Phonons and orbital degrees of freedom are likely to be
important in the context of the nematic order seem in the pnictide
superconductors, whose microscopic origin is not yet understood.

Of course, ultimately, the full signal is given by the contributions from all the relevant degrees of freedom.

\section{Outlook}
\label{sec:conclusions}

As the above examples have shown, the method presented here is
very powerful both in scope and predictive potential. We have, for
example, explicitly shown that various hidden orders could be
unambiguously identified. Moreover, as we tried to emphasize, this approach offers the advantage
of possibly helping with unbiased data analysis since all possible
contributions to the RIXS signal can in principle be systematically enumerated. 

With this theory in hand, where should one look next? As
proposed here, NiGa$_2$S$_4$ of course appears as a natural material to
investigate with RIXS or REXS. In particular, thanks to $S=1$
on Ni$^{2+}$ one expects ``direct RIXS'' processes to be involved and
therefore a strong signal. The current resolution on
RIXS instruments ---of about 130~meV--- is too low to
detect a sizable signal-to-noise ratio for a material where the exchange has been
estimated to lie at around $J\sim7$~meV (as boldly estimated from a Curie-Weiss temperature of
$|\Theta_{\rm CW}|\sim80$~K \cite{nakatsuji2005}). However, since
static order is expected (at higher
temperatures) \cite{stoudenmire2009}, Bragg peaks should
appear in REXS (see Fig.~\ref{fig:1}d)). Spin chain materials like LiCuVO$_4$ and
others \cite{nawa2014}, while perhaps even more promising in terms of
confidence in the realization of a nematic state, will have to await
the next generation of RIXS instruments, as their exchange energies are
also relatively low ($\sim30$~K). 
Perhaps, at this
point, high-quality data (like in the cuprates and iridates) would be worth
re-investigating in light of all the possibilities which our work
unearthed. One can, for example, imagine looking for signs of some of the ``stranger'' correlation
functions presented in Table~\ref{tab:couplings}. Another exciting
direction, briefly mentioned in Section~\ref{sec:other-degr-freed}, is
that of pnictide materials, as RIXS may help contribute to the effort of pinning down
the origin of the observed nematic order. Finally, most electrifying
would perhaps be the detection of chiral order in putative spin liquids on the
kagom\'{e} lattice \cite{hu2015,gong2015,wietek2015} or the possible
appearance of spin quadrupolar correlations (in the absence of dipolar ones) in La$_{2-x}$Ba$_x$Cu$_2$O$_4$, should it
display features of a spin density wave glass \cite{mross2015}.


With RIXS taking the central stage in various classes of systems, and new
resolution-improved machines on the horizon, the future seems bright
for refining our understanding of and discovering yet new physics in complex materials
amenable to RIXS. And with these general results {\sl and} derivation in this broad setting,
we hope to guide experiments as well as theory in this endeavor. It is
also our hope to have somewhat demystified the
understanding of RIXS for non-experts of microscopic calculations.

\acknowledgements

L.S.\ would like to thank Peter Armitage, Collin Broholm, Radu Coldea,
Natalia Drichko, David Hawthorn,
Bob Leheny, Kemp Plumb, Daniel Reich and especially Leon Balents for
useful discussions. L.S.\ was generously supported by the Gordon and
Betty Moore
Foundation through a postdoctoral fellowship of the EPiQS initiative,
Grant No.\ GBMF4303. T.S.\ was supported by NSF Grant DMR-1305741.  This work was also partially supported by a Simons Investigator award from the
Simons Foundation to Senthil Todadri.

\bibliography{rixs.bib}

\begin{thebibliography}{48}%
\makeatletter
\providecommand \@ifxundefined [1]{%
 \@ifx{#1\undefined}
}%
\providecommand \@ifnum [1]{%
 \ifnum #1\expandafter \@firstoftwo
 \else \expandafter \@secondoftwo
 \fi
}%
\providecommand \@ifx [1]{%
 \ifx #1\expandafter \@firstoftwo
 \else \expandafter \@secondoftwo
 \fi
}%
\providecommand \natexlab [1]{#1}%
\providecommand \enquote  [1]{``#1''}%
\providecommand \bibnamefont  [1]{#1}%
\providecommand \bibfnamefont [1]{#1}%
\providecommand \citenamefont [1]{#1}%
\providecommand \href@noop [0]{\@secondoftwo}%
\providecommand \href [0]{\begingroup \@sanitize@url \@href}%
\providecommand \@href[1]{\@@startlink{#1}\@@href}%
\providecommand \@@href[1]{\endgroup#1\@@endlink}%
\providecommand \@sanitize@url [0]{\catcode `\\12\catcode `\$12\catcode
  `\&12\catcode `\#12\catcode `\^12\catcode `\_12\catcode `\%12\relax}%
\providecommand \@@startlink[1]{}%
\providecommand \@@endlink[0]{}%
\providecommand \url  [0]{\begingroup\@sanitize@url \@url }%
\providecommand \@url [1]{\endgroup\@href {#1}{\urlprefix }}%
\providecommand \urlprefix  [0]{URL }%
\providecommand \Eprint [0]{\href }%
\providecommand \doibase [0]{http://dx.doi.org/}%
\providecommand \selectlanguage [0]{\@gobble}%
\providecommand \bibinfo  [0]{\@secondoftwo}%
\providecommand \bibfield  [0]{\@secondoftwo}%
\providecommand \translation [1]{[#1]}%
\providecommand \BibitemOpen [0]{}%
\providecommand \bibitemStop [0]{}%
\providecommand \bibitemNoStop [0]{.\EOS\space}%
\providecommand \EOS [0]{\spacefactor3000\relax}%
\providecommand \BibitemShut  [1]{\csname bibitem#1\endcsname}%
\let\auto@bib@innerbib\@empty
\bibitem [{\citenamefont {Penc}\ and\ \citenamefont
  {L\"{a}uchli}(2011)}]{penc2011}%
  \BibitemOpen
  \bibfield  {author} {\bibinfo {author} {\bibfnamefont {K.}~\bibnamefont
  {Penc}}\ and\ \bibinfo {author} {\bibfnamefont {A.~M.}\ \bibnamefont
  {L\"{a}uchli}},\ }\href {\doibase 10.1007/978-3-642-10589-0_13} {\emph
  {\bibinfo {title} {Introduction to Frustrated Magnetism}}},\ edited by\
  \bibinfo {editor} {\bibfnamefont {C.}~\bibnamefont {Lacroix}}, \bibinfo
  {editor} {\bibfnamefont {P.}~\bibnamefont {Mendels}}, \ and\ \bibinfo
  {editor} {\bibfnamefont {F.}~\bibnamefont {Mila}},\ \bibinfo {series}
  {Springer Series in Solid-State Sciences}, Vol.\ \bibinfo {volume} {164}\
  (\bibinfo  {publisher} {Springer Berlin Heidelberg},\ \bibinfo {year}
  {2011})\ pp.\ \bibinfo {pages} {331--362}\BibitemShut {NoStop}%
\bibitem [{\citenamefont {Chubukov}(1991)}]{chubukov1991}%
  \BibitemOpen
  \bibfield  {author} {\bibinfo {author} {\bibfnamefont {A.~V.}\ \bibnamefont
  {Chubukov}},\ }\bibfield  {title} {\enquote {\bibinfo {title} {Chiral,
  nematic, and dimer states in quantum spin chains},}\ }\href {\doibase
  10.1103/PhysRevB.44.4693} {\bibfield  {journal} {\bibinfo  {journal} {Phys.
  Rev. B}\ }\textbf {\bibinfo {volume} {44}},\ \bibinfo {pages} {4693--4696}
  (\bibinfo {year} {1991})}\BibitemShut {NoStop}%
\bibitem [{\citenamefont {Starykh}\ and\ \citenamefont
  {Balents}(2014)}]{starykh2014}%
  \BibitemOpen
  \bibfield  {author} {\bibinfo {author} {\bibfnamefont {O.~A.}\ \bibnamefont
  {Starykh}}\ and\ \bibinfo {author} {\bibfnamefont {L.}~\bibnamefont
  {Balents}},\ }\bibfield  {title} {\enquote {\bibinfo {title} {Excitations and
  quasi-one-dimensionality in field-induced nematic and spin density wave
  states},}\ }\href {\doibase 10.1103/PhysRevB.89.104407} {\bibfield  {journal}
  {\bibinfo  {journal} {Phys. Rev. B}\ }\textbf {\bibinfo {volume} {89}},\
  \bibinfo {pages} {104407} (\bibinfo {year} {2014})}\BibitemShut {NoStop}%
\bibitem [{\citenamefont {Messiah}(1962)}]{messiah1962}%
  \BibitemOpen
  \bibfield  {author} {\bibinfo {author} {\bibfnamefont {A.}~\bibnamefont
  {Messiah}},\ }\href@noop {} {\emph {\bibinfo {title} {Quantum Mechanics}}}\
  (\bibinfo {year} {1962})\BibitemShut {NoStop}%
\bibitem [{\citenamefont {Ament}\ \emph {et~al.}(2011)\citenamefont {Ament},
  \citenamefont {van Veenendaal}, \citenamefont {Devereaux}, \citenamefont
  {Hill},\ and\ \citenamefont {van~den Brink}}]{ament2011}%
  \BibitemOpen
  \bibfield  {author} {\bibinfo {author} {\bibfnamefont {L.~J.~P.}\
  \bibnamefont {Ament}}, \bibinfo {author} {\bibfnamefont {M.}~\bibnamefont
  {van Veenendaal}}, \bibinfo {author} {\bibfnamefont {T.~P.}\ \bibnamefont
  {Devereaux}}, \bibinfo {author} {\bibfnamefont {J.~P.}\ \bibnamefont {Hill}},
  \ and\ \bibinfo {author} {\bibfnamefont {J.}~\bibnamefont {van~den Brink}},\
  }\bibfield  {title} {\enquote {\bibinfo {title} {Resonant inelastic x-ray
  scattering studies of elementary excitations},}\ }\href {\doibase
  10.1103/RevModPhys.83.705} {\bibfield  {journal} {\bibinfo  {journal} {Rev.
  Mod. Phys.}\ }\textbf {\bibinfo {volume} {83}},\ \bibinfo {pages} {705--767}
  (\bibinfo {year} {2011})}\BibitemShut {NoStop}%
\bibitem [{\citenamefont {Dewey}\ \emph {et~al.}(1969)\citenamefont {Dewey},
  \citenamefont {Mapes},\ and\ \citenamefont {Reynolds}}]{dewey1969}%
  \BibitemOpen
  \bibfield  {author} {\bibinfo {author} {\bibfnamefont {R.~D.}\ \bibnamefont
  {Dewey}}, \bibinfo {author} {\bibfnamefont {R.~S.}\ \bibnamefont {Mapes}}, \
  and\ \bibinfo {author} {\bibfnamefont {T.~W.}\ \bibnamefont {Reynolds}},\
  }\href
  {http://www.kayelaby.npl.co.uk/atomic_and_nuclear_physics/4_2/4_2_1.html}
  {\emph {\bibinfo {title} {Handbook of X-ray and microprobe data: Tables of
  X-ray data}}},\ edited by\ \bibinfo {editor} {\bibnamefont {Elion}}\ and\
  \bibinfo {editor} {\bibnamefont {Stewart}}\ (\bibinfo  {publisher} {Oxford,
  Pergamon Press},\ \bibinfo {year} {1969})\BibitemShut {NoStop}%
\bibitem [{\citenamefont {Bearden}(1967)}]{bearden1967}%
  \BibitemOpen
  \bibfield  {author} {\bibinfo {author} {\bibfnamefont {J.~A.}\ \bibnamefont
  {Bearden}},\ }\bibfield  {title} {\enquote {\bibinfo {title} {X-ray
  wavelengths},}\ }\href {\doibase 10.1103/RevModPhys.39.78} {\bibfield
  {journal} {\bibinfo  {journal} {Rev. Mod. Phys.}\ }\textbf {\bibinfo {volume}
  {39}},\ \bibinfo {pages} {78--124} (\bibinfo {year} {1967})}\BibitemShut
  {NoStop}%
\bibitem [{\citenamefont {Ko}\ and\ \citenamefont {Lee}(2011)}]{ko2011}%
  \BibitemOpen
  \bibfield  {author} {\bibinfo {author} {\bibfnamefont {W.-H.}\ \bibnamefont
  {Ko}}\ and\ \bibinfo {author} {\bibfnamefont {P.~A.}\ \bibnamefont {Lee}},\
  }\bibfield  {title} {\enquote {\bibinfo {title} {Proposal for detecting
  spin-chirality terms in {Mott} insulators via resonant inelastic x-ray
  scattering},}\ }\href {\doibase 10.1103/PhysRevB.84.125102} {\bibfield
  {journal} {\bibinfo  {journal} {Phys. Rev. B}\ }\textbf {\bibinfo {volume}
  {84}},\ \bibinfo {pages} {125102} (\bibinfo {year} {2011})}\BibitemShut
  {NoStop}%
\bibitem [{\citenamefont {Michaud}\ \emph {et~al.}(2011)\citenamefont
  {Michaud}, \citenamefont {Vernay},\ and\ \citenamefont {Mila}}]{michaud2011}%
  \BibitemOpen
  \bibfield  {author} {\bibinfo {author} {\bibfnamefont {F.}~\bibnamefont
  {Michaud}}, \bibinfo {author} {\bibfnamefont {F.}~\bibnamefont {Vernay}}, \
  and\ \bibinfo {author} {\bibfnamefont {F.}~\bibnamefont {Mila}},\ }\bibfield
  {title} {\enquote {\bibinfo {title} {Theory of inelastic light scattering in
  spin-1 systems: Resonant regimes and detection of quadrupolar order},}\
  }\href {\doibase 10.1103/PhysRevB.84.184424} {\bibfield  {journal} {\bibinfo
  {journal} {Phys. Rev. B}\ }\textbf {\bibinfo {volume} {84}},\ \bibinfo
  {pages} {184424} (\bibinfo {year} {2011})}\BibitemShut {NoStop}%
\bibitem [{\citenamefont {Haverkort}(2010)}]{haverkort2010}%
  \BibitemOpen
  \bibfield  {author} {\bibinfo {author} {\bibfnamefont {M.~W.}\ \bibnamefont
  {Haverkort}},\ }\bibfield  {title} {\enquote {\bibinfo {title} {Theory of
  resonant inelastic x-ray scattering by collective magnetic excitations},}\
  }\href {\doibase 10.1103/PhysRevLett.105.167404} {\bibfield  {journal}
  {\bibinfo  {journal} {Phys. Rev. Lett.}\ }\textbf {\bibinfo {volume} {105}},\
  \bibinfo {pages} {167404} (\bibinfo {year} {2010})}\BibitemShut {NoStop}%
\bibitem [{Note1()}]{Note1}%
  \BibitemOpen
  \bibinfo {note} {The ``diamagnetic'' term $\protect \mathbf {A}^2$, of second
  order in $\protect \mathbf {A}$, is involved in the {\protect \sl
  first}-order contribution to the scattering amplitude, which is negligible
  close to resonance.}\BibitemShut {Stop}%
\bibitem [{Note2()}]{Note2}%
  \BibitemOpen
  \bibinfo {note} {For $\omega ^{\protect \rm x-ray}\sim 10$~keV, $|\protect
  \mathbf {k}|\sim 1$~\r A$^{-1}$ and $|\protect \mathbf {k}\cdot \delta
  \protect \mathbf {r}|\approx 0$ can seem hardly valid. In practice, however
  it has been shown to usually be a good approximation. Regardless, we discuss
  how to go beyond this approximation in Appendix~\ref
  {sec:higher-multipoles}.}\BibitemShut {Stop}%
\bibitem [{Note3()}]{Note3}%
  \BibitemOpen
  \bibinfo {note} {Indeed, $\tau \sim 10^{-15}$~s corresponds to an energy of
  order $4$~eV, which is typically that of a metal's bandwidth $W$. Taking
  estimate of an electron's velocity as $v=aW$ with $a$ the lattice spacing, we
  find that the travelled distance during time $\tau $ of order a lattice
  spacing.}\BibitemShut {Stop}%
\bibitem [{\citenamefont {Haverkort}\ \emph {et~al.}(2010)\citenamefont
  {Haverkort}, \citenamefont {Hollmann}, \citenamefont {Krug},\ and\
  \citenamefont {Tanaka}}]{haverkort2010b}%
  \BibitemOpen
  \bibfield  {author} {\bibinfo {author} {\bibfnamefont {M.~W.}\ \bibnamefont
  {Haverkort}}, \bibinfo {author} {\bibfnamefont {N.}~\bibnamefont {Hollmann}},
  \bibinfo {author} {\bibfnamefont {I.~P.}\ \bibnamefont {Krug}}, \ and\
  \bibinfo {author} {\bibfnamefont {A.}~\bibnamefont {Tanaka}},\ }\bibfield
  {title} {\enquote {\bibinfo {title} {Symmetry analysis of magneto-optical
  effects: The case of x-ray diffraction and x-ray absorption at the transition
  metal ${L}_{2,3}$ edge},}\ }\href {\doibase 10.1103/PhysRevB.82.094403}
  {\bibfield  {journal} {\bibinfo  {journal} {Phys. Rev. B}\ }\textbf {\bibinfo
  {volume} {82}},\ \bibinfo {pages} {094403} (\bibinfo {year}
  {2010})}\BibitemShut {NoStop}%
\bibitem [{Note4()}]{Note4}%
  \BibitemOpen
  \bibinfo {note} {More rigorously, one should derive the transition operators
  in terms of spin-orbit coupled effective spins, and {\protect \sl then}
  possibly neglect those which are not rotationally symmetric.}\BibitemShut
  {Stop}%
\bibitem [{\citenamefont {Snoke}()}]{snoke}%
  \BibitemOpen
  \bibfield  {author} {\bibinfo {author} {\bibfnamefont {D.~W.}\ \bibnamefont
  {Snoke}},\ }\href
  {http://www.phyast.pitt.edu/~snoke/resources/pointgroupshtml4thEd/PointGroupsMain4thEd.html}
  {\enquote {\bibinfo {title} {Point groups},}\ }\BibitemShut {NoStop}%
\bibitem [{Note5()}]{Note5}%
  \BibitemOpen
  \bibinfo {note} {Note that Ref.~\protect \rev@citealp {haverkort2010}
  additionally provides a relation between some of the coefficients $\alpha
  _\beta $ and absorption spectroscopy coefficients.}\BibitemShut {Stop}%
\bibitem [{\citenamefont {Nakatsuji}\ \emph {et~al.}(2005)\citenamefont
  {Nakatsuji}, \citenamefont {Nambu}, \citenamefont {Tonomura}, \citenamefont
  {Sakai}, \citenamefont {Jonas}, \citenamefont {Broholm}, \citenamefont
  {Tsunetsugu}, \citenamefont {Qiu},\ and\ \citenamefont
  {Maeno}}]{nakatsuji2005}%
  \BibitemOpen
  \bibfield  {author} {\bibinfo {author} {\bibfnamefont {S.}~\bibnamefont
  {Nakatsuji}}, \bibinfo {author} {\bibfnamefont {Y.}~\bibnamefont {Nambu}},
  \bibinfo {author} {\bibfnamefont {H.}~\bibnamefont {Tonomura}}, \bibinfo
  {author} {\bibfnamefont {O.}~\bibnamefont {Sakai}}, \bibinfo {author}
  {\bibfnamefont {S.}~\bibnamefont {Jonas}}, \bibinfo {author} {\bibfnamefont
  {C.}~\bibnamefont {Broholm}}, \bibinfo {author} {\bibfnamefont
  {H.}~\bibnamefont {Tsunetsugu}}, \bibinfo {author} {\bibfnamefont
  {Y.}~\bibnamefont {Qiu}}, \ and\ \bibinfo {author} {\bibfnamefont
  {Y.}~\bibnamefont {Maeno}},\ }\bibfield  {title} {\enquote {\bibinfo {title}
  {Spin disorder on a triangular lattice},}\ }\href {\doibase
  10.1126/science.1114727} {\bibfield  {journal} {\bibinfo  {journal}
  {Science}\ }\textbf {\bibinfo {volume} {309}},\ \bibinfo {pages} {1697--1700}
  (\bibinfo {year} {2005})},\ \Eprint
  {http://arxiv.org/abs/http://www.sciencemag.org/content/309/5741/1697.full.pdf}
  {http://www.sciencemag.org/content/309/5741/1697.full.pdf} \BibitemShut
  {NoStop}%
\bibitem [{\citenamefont {Tsunetsugu}\ and\ \citenamefont
  {Arikawa}(2006)}]{tsunetsugu2006}%
  \BibitemOpen
  \bibfield  {author} {\bibinfo {author} {\bibfnamefont {H.}~\bibnamefont
  {Tsunetsugu}}\ and\ \bibinfo {author} {\bibfnamefont {M.}~\bibnamefont
  {Arikawa}},\ }\bibfield  {title} {\enquote {\bibinfo {title} {Spin nematic
  phase in ${S}=1$ triangular antiferromagnets},}\ }\href {\doibase
  10.1143/JPSJ.75.083701} {\bibfield  {journal} {\bibinfo  {journal} {J. Phys.
  Soc. Jpn}\ }\textbf {\bibinfo {volume} {75}},\ \bibinfo {pages} {083701}
  (\bibinfo {year} {2006})},\ \Eprint
  {http://arxiv.org/abs/http://dx.doi.org/10.1143/JPSJ.75.083701}
  {http://dx.doi.org/10.1143/JPSJ.75.083701} \BibitemShut {NoStop}%
\bibitem [{\citenamefont {L\"{a}uchli}\ \emph {et~al.}(2006)\citenamefont
  {L\"{a}uchli}, \citenamefont {Mila},\ and\ \citenamefont
  {Penc}}]{lauchli2006}%
  \BibitemOpen
  \bibfield  {author} {\bibinfo {author} {\bibfnamefont {A.}~\bibnamefont
  {L\"{a}uchli}}, \bibinfo {author} {\bibfnamefont {F.}~\bibnamefont {Mila}}, \
  and\ \bibinfo {author} {\bibfnamefont {K.}~\bibnamefont {Penc}},\ }\bibfield
  {title} {\enquote {\bibinfo {title} {Quadrupolar phases of the ${S}=1$
  bilinear-biquadratic {Heisenberg} model on the triangular lattice},}\ }\href
  {\doibase 10.1103/PhysRevLett.97.087205} {\bibfield  {journal} {\bibinfo
  {journal} {Phys. Rev. Lett.}\ }\textbf {\bibinfo {volume} {97}},\ \bibinfo
  {pages} {087205} (\bibinfo {year} {2006})}\BibitemShut {NoStop}%
\bibitem [{\citenamefont {Bhattacharjee}\ \emph {et~al.}(2006)\citenamefont
  {Bhattacharjee}, \citenamefont {Shenoy},\ and\ \citenamefont
  {Senthil}}]{bhattacharjee2006}%
  \BibitemOpen
  \bibfield  {author} {\bibinfo {author} {\bibfnamefont {S.}~\bibnamefont
  {Bhattacharjee}}, \bibinfo {author} {\bibfnamefont {V.~B.}\ \bibnamefont
  {Shenoy}}, \ and\ \bibinfo {author} {\bibfnamefont {T.}~\bibnamefont
  {Senthil}},\ }\bibfield  {title} {\enquote {\bibinfo {title} {Possible
  ferro-spin nematic order in
  $\mathrm{Ni}{\mathrm{ga}}_{2}{\mathrm{s}}_{4}$},}\ }\href {\doibase
  10.1103/PhysRevB.74.092406} {\bibfield  {journal} {\bibinfo  {journal} {Phys.
  Rev. B}\ }\textbf {\bibinfo {volume} {74}},\ \bibinfo {pages} {092406}
  (\bibinfo {year} {2006})}\BibitemShut {NoStop}%
\bibitem [{\citenamefont {Stoudenmire}\ \emph {et~al.}(2009)\citenamefont
  {Stoudenmire}, \citenamefont {Trebst},\ and\ \citenamefont
  {Balents}}]{stoudenmire2009}%
  \BibitemOpen
  \bibfield  {author} {\bibinfo {author} {\bibfnamefont {E.~M.}\ \bibnamefont
  {Stoudenmire}}, \bibinfo {author} {\bibfnamefont {S.}~\bibnamefont {Trebst}},
  \ and\ \bibinfo {author} {\bibfnamefont {L.}~\bibnamefont {Balents}},\
  }\bibfield  {title} {\enquote {\bibinfo {title} {Quadrupolar correlations and
  spin freezing in ${S}=1$ triangular lattice antiferromagnets},}\ }\href
  {\doibase 10.1103/PhysRevB.79.214436} {\bibfield  {journal} {\bibinfo
  {journal} {Phys. Rev. B}\ }\textbf {\bibinfo {volume} {79}},\ \bibinfo
  {pages} {214436} (\bibinfo {year} {2009})}\BibitemShut {NoStop}%
\bibitem [{\citenamefont {Kaul}(2012)}]{kaul2012}%
  \BibitemOpen
  \bibfield  {author} {\bibinfo {author} {\bibfnamefont {R.}~\bibnamefont
  {Kaul}},\ }\bibfield  {title} {\enquote {\bibinfo {title} {Spin nematic
  ground state of the triangular lattice ${S}=1$ biquadratic model},}\ }\href
  {\doibase 10.1103/PhysRevB.86.104411} {\bibfield  {journal} {\bibinfo
  {journal} {Phys. Rev. B}\ }\textbf {\bibinfo {volume} {86}},\ \bibinfo
  {pages} {104411} (\bibinfo {year} {2012})}\BibitemShut {NoStop}%
\bibitem [{\citenamefont {Smerald}\ and\ \citenamefont
  {Shannon}(2013)}]{smerald2013}%
  \BibitemOpen
  \bibfield  {author} {\bibinfo {author} {\bibfnamefont {A.}~\bibnamefont
  {Smerald}}\ and\ \bibinfo {author} {\bibfnamefont {N.}~\bibnamefont
  {Shannon}},\ }\bibfield  {title} {\enquote {\bibinfo {title} {Theory of spin
  excitations in a quantum spin-nematic state},}\ }\href {\doibase
  10.1103/PhysRevB.88.184430} {\bibfield  {journal} {\bibinfo  {journal} {Phys.
  Rev. B}\ }\textbf {\bibinfo {volume} {88}},\ \bibinfo {pages} {184430}
  (\bibinfo {year} {2013})}\BibitemShut {NoStop}%
\bibitem [{\citenamefont {Pires}(2014)}]{pires2014}%
  \BibitemOpen
  \bibfield  {author} {\bibinfo {author} {\bibfnamefont {A.~S.~T.}\
  \bibnamefont {Pires}},\ }\bibfield  {title} {\enquote {\bibinfo {title}
  {Dynamics of the ferroquadrupolar phase of the ${S}=1$ bilinear-biquadratic
  model on the triangular lattice},}\ }\href {\doibase
  http://dx.doi.org/10.1016/j.ssc.2014.07.015} {\bibfield  {journal} {\bibinfo
  {journal} {Solid State Communications}\ }\textbf {\bibinfo {volume} {196}},\
  \bibinfo {pages} {24--27} (\bibinfo {year} {2014})}\BibitemShut {NoStop}%
\bibitem [{\citenamefont {V\"oll}\ and\ \citenamefont
  {Wessel}(2015)}]{voll2015}%
  \BibitemOpen
  \bibfield  {author} {\bibinfo {author} {\bibfnamefont {A.}~\bibnamefont
  {V\"oll}}\ and\ \bibinfo {author} {\bibfnamefont {S.}~\bibnamefont
  {Wessel}},\ }\bibfield  {title} {\enquote {\bibinfo {title} {Spin dynamics of
  the bilinear-biquadratic ${S}=1$ {Heisenberg} model on the triangular
  lattice: A quantum {Monte Carlo} study},}\ }\href {\doibase
  10.1103/PhysRevB.91.165128} {\bibfield  {journal} {\bibinfo  {journal} {Phys.
  Rev. B}\ }\textbf {\bibinfo {volume} {91}},\ \bibinfo {pages} {165128}
  (\bibinfo {year} {2015})}\BibitemShut {NoStop}%
\bibitem [{\citenamefont {Enderle}\ \emph {et~al.}(2005)\citenamefont
  {Enderle}, \citenamefont {Mukherjee}, \citenamefont {F\r{a}k}, \citenamefont
  {Kremer}, \citenamefont {Broto}, \citenamefont {Rosner}, \citenamefont
  {Drechsler}, \citenamefont {Richter}, \citenamefont {Malek}, \citenamefont
  {Prokofiev}, \citenamefont {Assmus}, \citenamefont {Pujol}, \citenamefont
  {Raggazzoni}, \citenamefont {Rakoto}, \citenamefont {Rheinst\"{a}dter},\ and\
  \citenamefont {R{\o}nnow}}]{enderle2005}%
  \BibitemOpen
  \bibfield  {author} {\bibinfo {author} {\bibfnamefont {M.}~\bibnamefont
  {Enderle}}, \bibinfo {author} {\bibfnamefont {C.}~\bibnamefont {Mukherjee}},
  \bibinfo {author} {\bibfnamefont {B.}~\bibnamefont {F\r{a}k}}, \bibinfo
  {author} {\bibfnamefont {R.~K.}\ \bibnamefont {Kremer}}, \bibinfo {author}
  {\bibfnamefont {J.-M.}\ \bibnamefont {Broto}}, \bibinfo {author}
  {\bibfnamefont {H.}~\bibnamefont {Rosner}}, \bibinfo {author} {\bibfnamefont
  {S.-L.}\ \bibnamefont {Drechsler}}, \bibinfo {author} {\bibfnamefont
  {J.}~\bibnamefont {Richter}}, \bibinfo {author} {\bibfnamefont
  {J.}~\bibnamefont {Malek}}, \bibinfo {author} {\bibfnamefont
  {A.}~\bibnamefont {Prokofiev}}, \bibinfo {author} {\bibfnamefont
  {W.}~\bibnamefont {Assmus}}, \bibinfo {author} {\bibfnamefont
  {S.}~\bibnamefont {Pujol}}, \bibinfo {author} {\bibfnamefont {J.-L.}\
  \bibnamefont {Raggazzoni}}, \bibinfo {author} {\bibfnamefont
  {H.}~\bibnamefont {Rakoto}}, \bibinfo {author} {\bibfnamefont
  {M.}~\bibnamefont {Rheinst\"{a}dter}}, \ and\ \bibinfo {author}
  {\bibfnamefont {H.~M.}\ \bibnamefont {R{\o}nnow}},\ }\bibfield  {title}
  {\enquote {\bibinfo {title} {Quantum helimagnetism of the frustrated spin-1/2
  chain {LiCuVO}$_4$},}\ }\href {\doibase 10.1209/epl/i2004-10484-x} {\bibfield
   {journal} {\bibinfo  {journal} {EPL (Europhysics Letters)}\ }\textbf
  {\bibinfo {volume} {70}},\ \bibinfo {pages} {237} (\bibinfo {year}
  {2005})}\BibitemShut {NoStop}%
\bibitem [{\citenamefont {Hagiwara}\ \emph {et~al.}(2011)\citenamefont
  {Hagiwara}, \citenamefont {Svistov}, \citenamefont {Fujita}, \citenamefont
  {Yamaguchi}, \citenamefont {Kimura}, \citenamefont {Omura}, \citenamefont
  {Prokofiev}, \citenamefont {Smirnov},\ and\ \citenamefont
  {Honda}}]{hagiwara2011}%
  \BibitemOpen
  \bibfield  {author} {\bibinfo {author} {\bibfnamefont {M.}~\bibnamefont
  {Hagiwara}}, \bibinfo {author} {\bibfnamefont {L.~E.}\ \bibnamefont
  {Svistov}}, \bibinfo {author} {\bibfnamefont {T.}~\bibnamefont {Fujita}},
  \bibinfo {author} {\bibfnamefont {H.}~\bibnamefont {Yamaguchi}}, \bibinfo
  {author} {\bibfnamefont {S.}~\bibnamefont {Kimura}}, \bibinfo {author}
  {\bibfnamefont {K.}~\bibnamefont {Omura}}, \bibinfo {author} {\bibfnamefont
  {A.}~\bibnamefont {Prokofiev}}, \bibinfo {author} {\bibfnamefont {A.~I.}\
  \bibnamefont {Smirnov}}, \ and\ \bibinfo {author} {\bibfnamefont
  {Z.}~\bibnamefont {Honda}},\ }\bibfield  {title} {\enquote {\bibinfo {title}
  {Possibility of the field-induced spin-nematic phase in {LiCuVO}$_4$},}\
  }\href {\doibase 10.1088/1742-6596/320/1/012049} {\bibfield  {journal}
  {\bibinfo  {journal} {Journal of Physics: Conference Series}\ }\textbf
  {\bibinfo {volume} {320}},\ \bibinfo {pages} {012049} (\bibinfo {year}
  {2011})}\BibitemShut {NoStop}%
\bibitem [{\citenamefont {Mourigal}\ \emph {et~al.}(2012)\citenamefont
  {Mourigal}, \citenamefont {Enderle}, \citenamefont {F\aa{}k}, \citenamefont
  {Kremer}, \citenamefont {Law}, \citenamefont {Schneidewind}, \citenamefont
  {Hiess},\ and\ \citenamefont {Prokofiev}}]{mourigal2012}%
  \BibitemOpen
  \bibfield  {author} {\bibinfo {author} {\bibfnamefont {M.}~\bibnamefont
  {Mourigal}}, \bibinfo {author} {\bibfnamefont {M.}~\bibnamefont {Enderle}},
  \bibinfo {author} {\bibfnamefont {B.}~\bibnamefont {F\aa{}k}}, \bibinfo
  {author} {\bibfnamefont {R.}~\bibnamefont {Kremer}}, \bibinfo {author}
  {\bibfnamefont {J.}~\bibnamefont {Law}}, \bibinfo {author} {\bibfnamefont
  {A.}~\bibnamefont {Schneidewind}}, \bibinfo {author} {\bibfnamefont
  {A.}~\bibnamefont {Hiess}}, \ and\ \bibinfo {author} {\bibfnamefont
  {A.}~\bibnamefont {Prokofiev}},\ }\bibfield  {title} {\enquote {\bibinfo
  {title} {Evidence of a bond-nematic phase in ${\mathrm{licuvo}}_{4}$},}\
  }\href {\doibase 10.1103/PhysRevLett.109.027203} {\bibfield  {journal}
  {\bibinfo  {journal} {Phys. Rev. Lett.}\ }\textbf {\bibinfo {volume} {109}},\
  \bibinfo {pages} {027203} (\bibinfo {year} {2012})}\BibitemShut {NoStop}%
\bibitem [{\citenamefont {Nawa}\ \emph {et~al.}(2014)\citenamefont {Nawa},
  \citenamefont {Okamoto}, \citenamefont {Matsuo}, \citenamefont {Kindo},
  \citenamefont {Kitahara}, \citenamefont {Yoshida}, \citenamefont {Ikeda},
  \citenamefont {Hara}, \citenamefont {Sakurai}, \citenamefont {Okubo},
  \citenamefont {Ohta},\ and\ \citenamefont {Hiroi}}]{nawa2014}%
  \BibitemOpen
  \bibfield  {author} {\bibinfo {author} {\bibfnamefont {K.}~\bibnamefont
  {Nawa}}, \bibinfo {author} {\bibfnamefont {Y.}~\bibnamefont {Okamoto}},
  \bibinfo {author} {\bibfnamefont {A.}~\bibnamefont {Matsuo}}, \bibinfo
  {author} {\bibfnamefont {K.}~\bibnamefont {Kindo}}, \bibinfo {author}
  {\bibfnamefont {Y.}~\bibnamefont {Kitahara}}, \bibinfo {author}
  {\bibfnamefont {S.a}\ \bibnamefont {Yoshida}}, \bibinfo {author}
  {\bibfnamefont {S.}~\bibnamefont {Ikeda}}, \bibinfo {author} {\bibfnamefont
  {S.}~\bibnamefont {Hara}}, \bibinfo {author} {\bibfnamefont {T.}~\bibnamefont
  {Sakurai}}, \bibinfo {author} {\bibfnamefont {S.}~\bibnamefont {Okubo}},
  \bibinfo {author} {\bibfnamefont {H.}~\bibnamefont {Ohta}}, \ and\ \bibinfo
  {author} {\bibfnamefont {Z.}~\bibnamefont {Hiroi}},\ }\bibfield  {title}
  {\enquote {\bibinfo {title} {{NaCuMoO}$_4${(OH)} as a candidate frustrated
  ${J}_1-{J}_2$ chain quantum magnet},}\ }\href {\doibase
  10.7566/JPSJ.83.103702} {\bibfield  {journal} {\bibinfo  {journal} {Journal
  of the Physical Society of Japan}\ }\textbf {\bibinfo {volume} {83}},\
  \bibinfo {pages} {103702} (\bibinfo {year} {2014})}\BibitemShut {NoStop}%
\bibitem [{\citenamefont {Kolezhuk}\ and\ \citenamefont
  {Vekua}(2005)}]{kolezhuzk2005}%
  \BibitemOpen
  \bibfield  {author} {\bibinfo {author} {\bibfnamefont {A.}~\bibnamefont
  {Kolezhuk}}\ and\ \bibinfo {author} {\bibfnamefont {T.}~\bibnamefont
  {Vekua}},\ }\bibfield  {title} {\enquote {\bibinfo {title} {Field-induced
  chiral phase in isotropic frustrated spin chains},}\ }\href {\doibase
  10.1103/PhysRevB.72.094424} {\bibfield  {journal} {\bibinfo  {journal} {Phys.
  Rev. B}\ }\textbf {\bibinfo {volume} {72}},\ \bibinfo {pages} {094424}
  (\bibinfo {year} {2005})}\BibitemShut {NoStop}%
\bibitem [{\citenamefont {Hikihara}\ \emph {et~al.}(2008)\citenamefont
  {Hikihara}, \citenamefont {Kecke}, \citenamefont {Momoi},\ and\ \citenamefont
  {Furusaki}}]{hikihara2008}%
  \BibitemOpen
  \bibfield  {author} {\bibinfo {author} {\bibfnamefont {T.}~\bibnamefont
  {Hikihara}}, \bibinfo {author} {\bibfnamefont {L.}~\bibnamefont {Kecke}},
  \bibinfo {author} {\bibfnamefont {T.}~\bibnamefont {Momoi}}, \ and\ \bibinfo
  {author} {\bibfnamefont {A.}~\bibnamefont {Furusaki}},\ }\bibfield  {title}
  {\enquote {\bibinfo {title} {Vector chiral and multipolar orders in the
  spin-$\frac{1}{2}$ frustrated ferromagnetic chain in magnetic field},}\
  }\href {\doibase 10.1103/PhysRevB.78.144404} {\bibfield  {journal} {\bibinfo
  {journal} {Phys. Rev. B}\ }\textbf {\bibinfo {volume} {78}},\ \bibinfo
  {pages} {144404} (\bibinfo {year} {2008})}\BibitemShut {NoStop}%
\bibitem [{\citenamefont {Sudan}\ \emph {et~al.}(2009)\citenamefont {Sudan},
  \citenamefont {L\"uscher},\ and\ \citenamefont {L\"auchli}}]{sudan2009}%
  \BibitemOpen
  \bibfield  {author} {\bibinfo {author} {\bibfnamefont {J.}~\bibnamefont
  {Sudan}}, \bibinfo {author} {\bibfnamefont {A.}~\bibnamefont {L\"uscher}}, \
  and\ \bibinfo {author} {\bibfnamefont {A.}~\bibnamefont {L\"auchli}},\
  }\bibfield  {title} {\enquote {\bibinfo {title} {Emergent multipolar spin
  correlations in a fluctuating spiral: The frustrated ferromagnetic
  spin-$\frac{1}{2}$ {Heisenberg} chain in a magnetic field},}\ }\href
  {\doibase 10.1103/PhysRevB.80.140402} {\bibfield  {journal} {\bibinfo
  {journal} {Phys. Rev. B}\ }\textbf {\bibinfo {volume} {80}},\ \bibinfo
  {pages} {140402} (\bibinfo {year} {2009})}\BibitemShut {NoStop}%
\bibitem [{\citenamefont {McCulloch}\ \emph {et~al.}(2008)\citenamefont
  {McCulloch}, \citenamefont {Kube}, \citenamefont {Kurz}, \citenamefont
  {Kleine}, \citenamefont {Schollw\"ock},\ and\ \citenamefont
  {Kolezhuk}}]{mcculloch2008}%
  \BibitemOpen
  \bibfield  {author} {\bibinfo {author} {\bibfnamefont {I.~P.}\ \bibnamefont
  {McCulloch}}, \bibinfo {author} {\bibfnamefont {R.}~\bibnamefont {Kube}},
  \bibinfo {author} {\bibfnamefont {M.}~\bibnamefont {Kurz}}, \bibinfo {author}
  {\bibfnamefont {A.}~\bibnamefont {Kleine}}, \bibinfo {author} {\bibfnamefont
  {U.}~\bibnamefont {Schollw\"ock}}, \ and\ \bibinfo {author} {\bibfnamefont
  {A.~K.}\ \bibnamefont {Kolezhuk}},\ }\bibfield  {title} {\enquote {\bibinfo
  {title} {Vector chiral order in frustrated spin chains},}\ }\href {\doibase
  10.1103/PhysRevB.77.094404} {\bibfield  {journal} {\bibinfo  {journal} {Phys.
  Rev. B}\ }\textbf {\bibinfo {volume} {77}},\ \bibinfo {pages} {094404}
  (\bibinfo {year} {2008})}\BibitemShut {NoStop}%
\bibitem [{\citenamefont {Marra}\ \emph {et~al.}(2013)\citenamefont {Marra},
  \citenamefont {Sykora}, \citenamefont {Wohlfeld},\ and\ \citenamefont
  {van~den Brink}}]{marra2013}%
  \BibitemOpen
  \bibfield  {author} {\bibinfo {author} {\bibfnamefont {P.}~\bibnamefont
  {Marra}}, \bibinfo {author} {\bibfnamefont {S.}~\bibnamefont {Sykora}},
  \bibinfo {author} {\bibfnamefont {K.}~\bibnamefont {Wohlfeld}}, \ and\
  \bibinfo {author} {\bibfnamefont {J.}~\bibnamefont {van~den Brink}},\
  }\bibfield  {title} {\enquote {\bibinfo {title} {Resonant inelastic x-ray
  scattering as a probe of the phase and excitations of the order parameter of
  superconductors},}\ }\href {\doibase 10.1103/PhysRevLett.110.117005}
  {\bibfield  {journal} {\bibinfo  {journal} {Phys. Rev. Lett.}\ }\textbf
  {\bibinfo {volume} {110}},\ \bibinfo {pages} {117005} (\bibinfo {year}
  {2013})}\BibitemShut {NoStop}%
\bibitem [{\citenamefont {Benjamin}\ \emph {et~al.}(2014)\citenamefont
  {Benjamin}, \citenamefont {Klich},\ and\ \citenamefont
  {Demler}}]{benjamin2014}%
  \BibitemOpen
  \bibfield  {author} {\bibinfo {author} {\bibfnamefont {D.}~\bibnamefont
  {Benjamin}}, \bibinfo {author} {\bibfnamefont {I.}~\bibnamefont {Klich}}, \
  and\ \bibinfo {author} {\bibfnamefont {E.}~\bibnamefont {Demler}},\
  }\bibfield  {title} {\enquote {\bibinfo {title} {Single-band model of
  resonant inelastic x-ray scattering by quasiparticles in high-${T}_{c}$
  cuprate superconductors},}\ }\href {\doibase 10.1103/PhysRevLett.112.247002}
  {\bibfield  {journal} {\bibinfo  {journal} {Phys. Rev. Lett.}\ }\textbf
  {\bibinfo {volume} {112}},\ \bibinfo {pages} {247002} (\bibinfo {year}
  {2014})}\BibitemShut {NoStop}%
\bibitem [{\citenamefont {Le~Tacon}\ \emph {et~al.}(2011)\citenamefont
  {Le~Tacon}, \citenamefont {Ghiringhelli}, \citenamefont {Chaloupka},
  \citenamefont {Sala}, \citenamefont {Hinkov}, \citenamefont {Haverkort},
  \citenamefont {Minola}, \citenamefont {Bakr}, \citenamefont {Zhou},
  \citenamefont {Blanco-Canosa}, \citenamefont {Monney}, \citenamefont {Song},
  \citenamefont {Sun}, \citenamefont {Lin}, \citenamefont {De~Luca},
  \citenamefont {Salluzzo}, \citenamefont {Khaliullin}, \citenamefont
  {Schmitt}, \citenamefont {Braicovich},\ and\ \citenamefont
  {Keimer}}]{letacon2011}%
  \BibitemOpen
  \bibfield  {author} {\bibinfo {author} {\bibfnamefont {M.}~\bibnamefont
  {Le~Tacon}}, \bibinfo {author} {\bibfnamefont {G.}~\bibnamefont
  {Ghiringhelli}}, \bibinfo {author} {\bibfnamefont {J.}~\bibnamefont
  {Chaloupka}}, \bibinfo {author} {\bibfnamefont {M.~Moretti}\ \bibnamefont
  {Sala}}, \bibinfo {author} {\bibfnamefont {V.}~\bibnamefont {Hinkov}},
  \bibinfo {author} {\bibfnamefont {M.~W.}\ \bibnamefont {Haverkort}}, \bibinfo
  {author} {\bibfnamefont {M.}~\bibnamefont {Minola}}, \bibinfo {author}
  {\bibfnamefont {M.}~\bibnamefont {Bakr}}, \bibinfo {author} {\bibfnamefont
  {K.~J.}\ \bibnamefont {Zhou}}, \bibinfo {author} {\bibfnamefont
  {S.}~\bibnamefont {Blanco-Canosa}}, \bibinfo {author} {\bibfnamefont
  {C.}~\bibnamefont {Monney}}, \bibinfo {author} {\bibfnamefont {Y.~T.}\
  \bibnamefont {Song}}, \bibinfo {author} {\bibfnamefont {G.~L.}\ \bibnamefont
  {Sun}}, \bibinfo {author} {\bibfnamefont {C.~T.}\ \bibnamefont {Lin}},
  \bibinfo {author} {\bibfnamefont {G.~M.}\ \bibnamefont {De~Luca}}, \bibinfo
  {author} {\bibfnamefont {M.}~\bibnamefont {Salluzzo}}, \bibinfo {author}
  {\bibfnamefont {G.}~\bibnamefont {Khaliullin}}, \bibinfo {author}
  {\bibfnamefont {T.}~\bibnamefont {Schmitt}}, \bibinfo {author} {\bibfnamefont
  {L.}~\bibnamefont {Braicovich}}, \ and\ \bibinfo {author} {\bibfnamefont
  {B.}~\bibnamefont {Keimer}},\ }\bibfield  {title} {\enquote {\bibinfo {title}
  {Intense paramagnon excitations in a large family of high-temperature
  superconductors},}\ }\href {\doibase 10.1038/nphys2041} {\bibfield  {journal}
  {\bibinfo  {journal} {Nat. Phys.}\ }\textbf {\bibinfo {volume} {7}},\
  \bibinfo {pages} {725--730} (\bibinfo {year} {2011})}\BibitemShut {NoStop}%
\bibitem [{\citenamefont {Le~Tacon}\ \emph {et~al.}(2013)\citenamefont
  {Le~Tacon}, \citenamefont {Minola}, \citenamefont {Peets}, \citenamefont
  {Moretti~Sala}, \citenamefont {Blanco-Canosa}, \citenamefont {Hinkov},
  \citenamefont {Liang}, \citenamefont {Bonn}, \citenamefont {Hardy},
  \citenamefont {Lin}, \citenamefont {Schmitt}, \citenamefont {Braicovich},
  \citenamefont {Ghiringhelli},\ and\ \citenamefont {Keimer}}]{letacon2013}%
  \BibitemOpen
  \bibfield  {author} {\bibinfo {author} {\bibfnamefont {M.}~\bibnamefont
  {Le~Tacon}}, \bibinfo {author} {\bibfnamefont {M.}~\bibnamefont {Minola}},
  \bibinfo {author} {\bibfnamefont {D.~C.}\ \bibnamefont {Peets}}, \bibinfo
  {author} {\bibfnamefont {M.}~\bibnamefont {Moretti~Sala}}, \bibinfo {author}
  {\bibfnamefont {S.}~\bibnamefont {Blanco-Canosa}}, \bibinfo {author}
  {\bibfnamefont {V.}~\bibnamefont {Hinkov}}, \bibinfo {author} {\bibfnamefont
  {R.}~\bibnamefont {Liang}}, \bibinfo {author} {\bibfnamefont {D.~A.}\
  \bibnamefont {Bonn}}, \bibinfo {author} {\bibfnamefont {W.~N.}\ \bibnamefont
  {Hardy}}, \bibinfo {author} {\bibfnamefont {C.~T.}\ \bibnamefont {Lin}},
  \bibinfo {author} {\bibfnamefont {T.}~\bibnamefont {Schmitt}}, \bibinfo
  {author} {\bibfnamefont {L.}~\bibnamefont {Braicovich}}, \bibinfo {author}
  {\bibfnamefont {G.}~\bibnamefont {Ghiringhelli}}, \ and\ \bibinfo {author}
  {\bibfnamefont {B.}~\bibnamefont {Keimer}},\ }\bibfield  {title} {\enquote
  {\bibinfo {title} {Dispersive spin excitations in highly overdoped cuprates
  revealed by resonant inelastic x-ray scattering},}\ }\href {\doibase
  10.1103/PhysRevB.88.020501} {\bibfield  {journal} {\bibinfo  {journal} {Phys.
  Rev. B}\ }\textbf {\bibinfo {volume} {88}},\ \bibinfo {pages} {020501}
  (\bibinfo {year} {2013})}\BibitemShut {NoStop}%
\bibitem [{\citenamefont {Dean}\ \emph {et~al.}(2013)\citenamefont {Dean},
  \citenamefont {Dellea}, \citenamefont {Springell}, \citenamefont
  {Yakhou-Harris}, \citenamefont {Kummer}, \citenamefont {Brookes},
  \citenamefont {Liu}, \citenamefont {Sun}, \citenamefont {Strle},
  \citenamefont {Schmitt}, \citenamefont {Braicovich}, \citenamefont
  {Ghiringhelli}, \citenamefont {Bozovic},\ and\ \citenamefont
  {Hill}}]{dean2013}%
  \BibitemOpen
  \bibfield  {author} {\bibinfo {author} {\bibfnamefont {M.~P.~M.}\
  \bibnamefont {Dean}}, \bibinfo {author} {\bibfnamefont {G.}~\bibnamefont
  {Dellea}}, \bibinfo {author} {\bibfnamefont {R.~S.}\ \bibnamefont
  {Springell}}, \bibinfo {author} {\bibfnamefont {F.}~\bibnamefont
  {Yakhou-Harris}}, \bibinfo {author} {\bibfnamefont {K.}~\bibnamefont
  {Kummer}}, \bibinfo {author} {\bibfnamefont {N.~B.}\ \bibnamefont {Brookes}},
  \bibinfo {author} {\bibfnamefont {X.}~\bibnamefont {Liu}}, \bibinfo {author}
  {\bibfnamefont {Y-J.}\ \bibnamefont {Sun}}, \bibinfo {author} {\bibfnamefont
  {J.}~\bibnamefont {Strle}}, \bibinfo {author} {\bibfnamefont
  {T.}~\bibnamefont {Schmitt}}, \bibinfo {author} {\bibfnamefont
  {L.}~\bibnamefont {Braicovich}}, \bibinfo {author} {\bibfnamefont
  {G.}~\bibnamefont {Ghiringhelli}}, \bibinfo {author} {\bibfnamefont
  {I.}~\bibnamefont {Bozovic}}, \ and\ \bibinfo {author} {\bibfnamefont
  {J.~P.}\ \bibnamefont {Hill}},\ }\bibfield  {title} {\enquote {\bibinfo
  {title} {Persistence of magnetic excitations in {La}$_{2-x}${Sr}$_x${CuO}$_4$
  from the undoped insulator to the heavily overdoped non-superconducting
  metal},}\ }\href {\doibase 10.1038/nmat3723} {\bibfield  {journal} {\bibinfo
  {journal} {Nat. Mater.}\ }\textbf {\bibinfo {volume} {12}},\ \bibinfo {pages}
  {1019--1023} (\bibinfo {year} {2013})}\BibitemShut {NoStop}%
\bibitem [{\citenamefont {Wakimoto}\ \emph {et~al.}(2015)\citenamefont
  {Wakimoto}, \citenamefont {Ishii}, \citenamefont {Kimura}, \citenamefont
  {Fujita}, \citenamefont {Dellea}, \citenamefont {Kummer}, \citenamefont
  {Braicovich}, \citenamefont {Ghiringhelli}, \citenamefont {Debeer-Schmitt},\
  and\ \citenamefont {Granroth}}]{wakimoto2015}%
  \BibitemOpen
  \bibfield  {author} {\bibinfo {author} {\bibfnamefont {S.}~\bibnamefont
  {Wakimoto}}, \bibinfo {author} {\bibfnamefont {K.}~\bibnamefont {Ishii}},
  \bibinfo {author} {\bibfnamefont {H.}~\bibnamefont {Kimura}}, \bibinfo
  {author} {\bibfnamefont {M.}~\bibnamefont {Fujita}}, \bibinfo {author}
  {\bibfnamefont {G.}~\bibnamefont {Dellea}}, \bibinfo {author} {\bibfnamefont
  {K.}~\bibnamefont {Kummer}}, \bibinfo {author} {\bibfnamefont
  {L.}~\bibnamefont {Braicovich}}, \bibinfo {author} {\bibfnamefont
  {G.}~\bibnamefont {Ghiringhelli}}, \bibinfo {author} {\bibfnamefont {L.~M.}\
  \bibnamefont {Debeer-Schmitt}}, \ and\ \bibinfo {author} {\bibfnamefont
  {G.~E.}\ \bibnamefont {Granroth}},\ }\bibfield  {title} {\enquote {\bibinfo
  {title} {High-energy magnetic excitations in overdoped
  {La}$_{2−x}${Sr}$_x${CuO}$_4$ studied by neutron and resonant inelastic
  x-ray scattering},}\ }\href {http://arxiv.org/abs/1505.03945} {\bibfield
  {journal} {\bibinfo  {journal} {arXiv:1505.03945}\ } (\bibinfo {year}
  {2015})}\BibitemShut {NoStop}%
\bibitem [{\citenamefont {Hu}\ \emph {et~al.}(2015)\citenamefont {Hu},
  \citenamefont {Zhu}, \citenamefont {Zhang}, \citenamefont {Gong},
  \citenamefont {Becca},\ and\ \citenamefont {Sheng}}]{hu2015}%
  \BibitemOpen
  \bibfield  {author} {\bibinfo {author} {\bibfnamefont {W.-J.}\ \bibnamefont
  {Hu}}, \bibinfo {author} {\bibfnamefont {W.}~\bibnamefont {Zhu}}, \bibinfo
  {author} {\bibfnamefont {Y.}~\bibnamefont {Zhang}}, \bibinfo {author}
  {\bibfnamefont {S.}~\bibnamefont {Gong}}, \bibinfo {author} {\bibfnamefont
  {F.}~\bibnamefont {Becca}}, \ and\ \bibinfo {author} {\bibfnamefont {D.~N.}\
  \bibnamefont {Sheng}},\ }\bibfield  {title} {\enquote {\bibinfo {title}
  {Variational monte carlo study of a chiral spin liquid in the extended
  {Heisenberg} model on the kagome lattice},}\ }\href {\doibase
  10.1103/PhysRevB.91.041124} {\bibfield  {journal} {\bibinfo  {journal} {Phys.
  Rev. B}\ }\textbf {\bibinfo {volume} {91}},\ \bibinfo {pages} {041124}
  (\bibinfo {year} {2015})}\BibitemShut {NoStop}%
\bibitem [{\citenamefont {Gong}\ \emph {et~al.}(2015)\citenamefont {Gong},
  \citenamefont {Zhu}, \citenamefont {Balents},\ and\ \citenamefont
  {Sheng}}]{gong2015}%
  \BibitemOpen
  \bibfield  {author} {\bibinfo {author} {\bibfnamefont {S.-S.}\ \bibnamefont
  {Gong}}, \bibinfo {author} {\bibfnamefont {W.}~\bibnamefont {Zhu}}, \bibinfo
  {author} {\bibfnamefont {L.}~\bibnamefont {Balents}}, \ and\ \bibinfo
  {author} {\bibfnamefont {D.~N.}\ \bibnamefont {Sheng}},\ }\bibfield  {title}
  {\enquote {\bibinfo {title} {Global phase diagram of competing ordered and
  quantum spin-liquid phases on the kagome lattice},}\ }\href {\doibase
  10.1103/PhysRevB.91.075112} {\bibfield  {journal} {\bibinfo  {journal} {Phys.
  Rev. B}\ }\textbf {\bibinfo {volume} {91}},\ \bibinfo {pages} {075112}
  (\bibinfo {year} {2015})}\BibitemShut {NoStop}%
\bibitem [{\citenamefont {Wietek}\ \emph {et~al.}(2015)\citenamefont {Wietek},
  \citenamefont {Sterdyniak},\ and\ \citenamefont {L\"{a}uchli}}]{wietek2015}%
  \BibitemOpen
  \bibfield  {author} {\bibinfo {author} {\bibfnamefont {A.}~\bibnamefont
  {Wietek}}, \bibinfo {author} {\bibfnamefont {A.}~\bibnamefont {Sterdyniak}},
  \ and\ \bibinfo {author} {\bibfnamefont {A.~M.}\ \bibnamefont
  {L\"{a}uchli}},\ }\bibfield  {title} {\enquote {\bibinfo {title} {Nature of
  chiral spin liquids on the kagome lattice},}\ }\href
  {http://arxiv.org/abs/1503.03389} {\  (\bibinfo {year} {2015})},\ \Eprint
  {http://arxiv.org/abs/arXiv:1503.03389} {arXiv:1503.03389} \BibitemShut
  {NoStop}%
\bibitem [{\citenamefont {Mross}\ and\ \citenamefont
  {Senthil}(2015)}]{mross2015}%
  \BibitemOpen
  \bibfield  {author} {\bibinfo {author} {\bibfnamefont {D.~F.}\ \bibnamefont
  {Mross}}\ and\ \bibinfo {author} {\bibfnamefont {T.}~\bibnamefont
  {Senthil}},\ }\bibfield  {title} {\enquote {\bibinfo {title} {Spin and pair
  density wave glasses},}\ }\href {http://arxiv.org/abs/1502.00002} {\
  (\bibinfo {year} {2015})},\ \Eprint {http://arxiv.org/abs/arXiv:1502.00002}
  {arXiv:1502.00002} \BibitemShut {NoStop}%
\bibitem [{Note6()}]{Note6}%
  \BibitemOpen
  \bibinfo {note} {One may also write the $R$ transformation as $S^\mu
  \rightarrow U^\dagger _R S^\mu U_R$, where $U_R$ acts in spin space ($U_R$ is
  the operator that $R$ maps into through the appropriate representation of the
  symmetry group).}\BibitemShut {Stop}%
\bibitem [{\citenamefont {Bell}(1954)}]{bell1954}%
  \BibitemOpen
  \bibfield  {author} {\bibinfo {author} {\bibfnamefont {Dorothy~G.}\
  \bibnamefont {Bell}},\ }\bibfield  {title} {\enquote {\bibinfo {title} {Group
  theory and crystal lattices},}\ }\href {\doibase 10.1103/RevModPhys.26.311}
  {\bibfield  {journal} {\bibinfo  {journal} {Rev. Mod. Phys.}\ }\textbf
  {\bibinfo {volume} {26}},\ \bibinfo {pages} {311--320} (\bibinfo {year}
  {1954})}\BibitemShut {NoStop}%
\bibitem [{\citenamefont {Koster}\ \emph {et~al.}(1963)\citenamefont {Koster},
  \citenamefont {Dimmock}, \citenamefont {Wheeler},\ and\ \citenamefont
  {Statz}}]{koster1963}%
  \BibitemOpen
  \bibfield  {author} {\bibinfo {author} {\bibfnamefont {G.~F.}\ \bibnamefont
  {Koster}}, \bibinfo {author} {\bibfnamefont {J.~O.}\ \bibnamefont {Dimmock}},
  \bibinfo {author} {\bibfnamefont {R.~G.}\ \bibnamefont {Wheeler}}, \ and\
  \bibinfo {author} {\bibfnamefont {H.}~\bibnamefont {Statz}},\ }\href
  {http://www.phyast.pitt.edu/~snoke/resources/pointgroupshtml4thEd/PointGroupsMain4thEd.html}
  {\emph {\bibinfo {title} {Properties of the Thirty-Two Point Groups}}}\
  (\bibinfo  {publisher} {MIT Press, Cambridge},\ \bibinfo {year}
  {1963})\BibitemShut {NoStop}%
\bibitem [{\citenamefont {Goss}()}]{goss}%
  \BibitemOpen
  \bibfield  {author} {\bibinfo {author} {\bibfnamefont {J.}~\bibnamefont
  {Goss}},\ }\href {https://www.staff.ncl.ac.uk/j.p.goss/symmetry/} {\enquote
  {\bibinfo {title} {Point group symmetry},}\ }\BibitemShut {NoStop}%
\end{thebibliography}%

\appendix

\section{Electromagnetic field}
\label{sec:electr-field}

As mentioned in the main text, the electromagnetic vector potential at point $\mathbf{r}$ may be expanded in plane waves 
\begin{equation}
  \label{eq:2}
  \mathbf{A}(\mathbf{r})=\sum_\mathbf{k}\sqrt{\frac{\hbar}{2V\epsilon_0\omega_\mathbf{k}}}\sum_{{\boldsymbol{\varepsilon}}}\left({\boldsymbol{\varepsilon}}^*a_{\mathbf{k},{\boldsymbol{\varepsilon}}}^\dagger
    e^{-i\mathbf{k}\cdot\mathbf{r}}+{\rm
    h.c.}\right),
\end{equation}
where $\hbar$ is Planck's constant, $\epsilon_0$ is the vacuum
dielectric polarization,
$\omega_\mathbf{k}=\omega_{-\mathbf{k}}=c|\mathbf{k}|$, with $c$ the
speed of light, $V$ is the volume in which the electromagnetic
field is confined, ${\boldsymbol{\varepsilon}}$ has nonzero
components only along (real) vectors perpendicular to
$\mathbf{k}$, and we define $
\mathbf{A}_{\mathbf{k},{\boldsymbol{\varepsilon}}}(\mathbf{r})={\boldsymbol{\varepsilon}}^*a_{\mathbf{k},{\boldsymbol{\varepsilon}}}^\dagger
    e^{-i\mathbf{k}\cdot\mathbf{r}}+{\rm
    h.c.}$. $a_{\mathbf{k},{\boldsymbol{\varepsilon}}}^\dagger$
is the creation operator of a photon of momentum $\mathbf{k}$ and
polarization (helicity) ${\boldsymbol{\varepsilon}}$. This ``expansion'' introduces (and defines) the
polarization vector ${\boldsymbol{\varepsilon}}$ which encodes the
vectorial (in the sense of a tensor of rank one) nature of the $S=1$ field
$\mathbf{A}$. We return to the symmetry transformation rules of
$\mathbf{A}$ and ${\boldsymbol{\varepsilon}}$ in Appendix~\ref{sec:transformation-rules}.

\section{Transformation rules}
\label{sec:transformation-rules}

The (pseudo-)vector of spin operators $\mathbf{S}_\mathbf{r}$ transforms under
a spatial operation $R$ and time reversal (TR)
according to 
\begin{equation}
  \label{eq:4}
  \begin{cases}
    R:&\mathbf{S}_\mathbf{r}\rightarrow
    \det(R)R\cdot\mathbf{S}_{R\cdot\mathbf{r}}\\
{\rm TR}:&\mathbf{S}_\mathbf{r}\rightarrow
    -\mathbf{S}_{\mathbf{r}}
  \end{cases},
\end{equation}
regardless of the value of $S(S+1)$ \footnote{One may also write the
  $R$ transformation as $S^\mu\rightarrow U^\dagger_R S^\mu U_R$,
  where $U_R$ acts in spin space ($U_R$ is the operator that $R$ maps into through the
  appropriate representation of the symmetry group).}. The vector potential $\mathbf{A}$ transforms as
\begin{equation}
  \label{eq:5}
  \begin{cases}
        R:&\mathbf{A}(\mathbf{r})\rightarrow
    R\cdot\mathbf{A}({R\cdot\mathbf{r}})\\
{\rm TR}:&\mathbf{A}(\mathbf{r})\rightarrow
    -\mathbf{A}({\mathbf{r}})
  \end{cases},
\end{equation}
so that the polarization ${\boldsymbol{\varepsilon}}$ transforms
according to 
\begin{equation}
  \label{eq:6}
    \begin{cases}
        R:&{\boldsymbol{\varepsilon}}\rightarrow
    R\cdot{\boldsymbol{\varepsilon}}\\
{\rm TR}:&{\boldsymbol{\varepsilon}}\rightarrow
    -{\boldsymbol{\varepsilon}}^*
  \end{cases}.
\end{equation}
Note that the definition of the polarization sometimes differs by,
e.g., a factor of $i$, and the polarization is then ``even'' (times complex conjugation) under the time
reversal operation.

If the spatial symmetry group contains all spherical operations (which
contain in particular all $SO(3)$ operations),
$\mathbf{S}$ and ${\boldsymbol{\varepsilon}}$ transform under the
``$L=1$'' representation of $SO(3)$ (regardless of the value of
$S(S+1)$). Note that here, the name ``$L$'' is purely formal. Using
the notation from Ref.~\onlinecite{snoke} for the full rotational
symmetry group ``$D$'' ($SO(3)\subset D$), ${\boldsymbol{\varepsilon}}$ and $\mathbf{S}$ transform under
the $D_1^-$ and $D_1^+$ representations, respectively, where $\pm$
indicate parity under the inversion transformation.


\section{Derivation of Table I}
\label{sec:derivation-table-i}

In the equations below, the numbers are representation labels
($L=0,1,2,...$ associated to $D_L^\pm$), and the
superscripts {\sl schematically} show basis elements (in the form of tensors) in terms of the original terms in the
products. Products of representations for
\begin{itemize}
\item zero spins:
\begin{equation}
1^{{\boldsymbol{\varepsilon}}'}\times
  1^{{\boldsymbol{\varepsilon}}}
=\left(0^{{\boldsymbol{\varepsilon}}'\cdot
    {{\boldsymbol{\varepsilon}}}}+1^{{\boldsymbol{\varepsilon}}'\times
    {{\boldsymbol{\varepsilon}}}}+2^{\llbracket{\boldsymbol{\varepsilon}}',
    {{\boldsymbol{\varepsilon}}}\rrbracket}\right);
\end{equation}
\item one spin:
\begin{equation}
1^{{\boldsymbol{\varepsilon}}'}\times
  1^{{\boldsymbol{\varepsilon}}}\times1^{\mathbf{S}_i}
=\left(0^{{\boldsymbol{\varepsilon}}'\cdot
    {{\boldsymbol{\varepsilon}}}}+1^{{\boldsymbol{\varepsilon}}'\times
    {{\boldsymbol{\varepsilon}}}}+2^{\llbracket{\boldsymbol{\varepsilon}}',
    {{\boldsymbol{\varepsilon}}}\rrbracket}\right)\times
1^{\mathbf{S}_i};
\end{equation}
\item two spins:
\begin{eqnarray}
  1^{{\boldsymbol{\varepsilon}}'}\times
  1^{{\boldsymbol{\varepsilon}}}\times1^{\mathbf{S}_i}\times1^{\mathbf{S}_j}&=&\left(0^{{\boldsymbol{\varepsilon}}'\cdot
    {{\boldsymbol{\varepsilon}}}}+1^{{\boldsymbol{\varepsilon}}'\times
    {{\boldsymbol{\varepsilon}}}}+2^{\llbracket{\boldsymbol{\varepsilon}}',
    {{\boldsymbol{\varepsilon}}}\rrbracket}\right)\\
&&\quad\times
\left(0^{\mathbf{S}_i\cdot\mathbf{S}_j}+1^{\mathbf{S}_i\times\mathbf{S}_j}+2^{\llbracket\mathbf{S}_i,\mathbf{S}_j\rrbracket}\right);\nonumber
\end{eqnarray}
\item three spins:
\begin{eqnarray}
&&  1^{{\boldsymbol{\varepsilon}}'}\times
  1^{{\boldsymbol{\varepsilon}}}\times1^{\mathbf{S}_i}\times1^{\mathbf{S}_j}\times1^{\mathbf{S}_k}\\
&=&\left(0^{{\boldsymbol{\varepsilon}}'\cdot
    {{\boldsymbol{\varepsilon}}}}+1^{{\boldsymbol{\varepsilon}}'\times
    {{\boldsymbol{\varepsilon}}}}+2^{\llbracket{\boldsymbol{\varepsilon}}',
    {{\boldsymbol{\varepsilon}}}\rrbracket}\right)\nonumber\\
&&\qquad\qquad\times
\left(0^{\mathbf{S}_i\cdot\mathbf{S}_j}+1^{\mathbf{S}_i\times\mathbf{S}_j}+2^{\llbracket\mathbf{S}_i,\mathbf{S}_j\rrbracket}\right)\times1^{\mathbf{S}_k}\nonumber\\
&=&\left(0^{{\boldsymbol{\varepsilon}}'\cdot
    {{\boldsymbol{\varepsilon}}}}+1^{{\boldsymbol{\varepsilon}}'\times
    {{\boldsymbol{\varepsilon}}}}+2^{\llbracket{\boldsymbol{\varepsilon}}',
    {{\boldsymbol{\varepsilon}}}\rrbracket}\right)\nonumber\\
&&\times
\left(1^{(\mathbf{S}_i\cdot\mathbf{S}_j)\mathbf{S}_k}+0^{(\mathbf{S}_i\times\mathbf{S}_j)\cdot\mathbf{S}_k}+1^{(\mathbf{S}_i\times\mathbf{S}_j)\times\mathbf{S}_k}\right.\nonumber\\
&&\qquad\qquad+2^{\llbracket(\mathbf{S}_i\times\mathbf{S}_j),\mathbf{S}_k\rrbracket}+1^{\llbracket\mathbf{S}_i,\mathbf{S}_j\rrbracket\cdot\mathbf{S}_k}+2^{\llbracket\mathbf{S}_i,\mathbf{S}_j\rrbracket\times\mathbf{S}_k}\nonumber\\
&&\left.\qquad\qquad\qquad\qquad+3^{\llbracket\llbracket\mathbf{S}_i,\mathbf{S}_j\rrbracket,\mathbf{S}_k\rrbracket}\right),\nonumber
\end{eqnarray}
\end{itemize}
where the definition of the double brackets has been extended to:
\begin{equation}
  \label{eq:18}
\begin{cases}
  (\llbracket\mathbf{u},\mathbf{v}\rrbracket\cdot\mathbf{w})_\mu=\sum_\nu\llbracket\mathbf{u},\mathbf{v}\rrbracket_{\mu\nu}w_\nu\\
  (\llbracket\mathbf{u},\mathbf{v}\rrbracket\times\mathbf{w})_{\mu\rho}=\sum_{\nu,\lambda}\epsilon_{\nu\lambda\rho}\llbracket\mathbf{u},\mathbf{v}\rrbracket_{\mu\nu}w_\lambda\\
 (\llbracket\llbracket\mathbf{u},\mathbf{v}\rrbracket,\mathbf{w}\rrbracket)_{\mu\nu\lambda}=\llbracket\mathbf{u},\mathbf{v}\rrbracket_{\mu\nu}w_\lambda
\end{cases}
\end{equation}
Only products of terms belonging to the same representation will
have a contribution in the ``final'' $0$ representation (by
contracting all the indices).

Explicitly, the operator obtained for all the terms in
Table~\ref{tab:couplings} reads:
\begin{eqnarray}
  \label{eq:13}
&&
T=\left({\boldsymbol{\varepsilon}}'{}^*\cdot{\boldsymbol{\varepsilon}}\right)\left[a_{0,1}\mathbf{S}_i\cdot\mathbf{S}_j+a_{0,2}(\mathbf{S}_i\times\mathbf{S}_j)\cdot\mathbf{S}_k\right]\\
&&\quad+
\left({\boldsymbol{\varepsilon}}'{}^*\times{\boldsymbol{\varepsilon}}\right)\cdot\left[a_{1,1}\mathbf{S}_i+a_{1,2}\mathbf{S}_i\times\mathbf{S}_j+a_{1,3}\left(\mathbf{S}_i\cdot\mathbf{S}_j\right)\mathbf{S}_k\right.\nonumber\\
&&\left.\qquad\qquad\qquad+a_{1,4}\left(\mathbf{S}_i\times\mathbf{S}_j\right)\times\mathbf{S}_k+a_{1,5}\left\llbracket\mathbf{S}_i,\mathbf{S}_j\right\rrbracket\cdot\mathbf{S}_k\right]\nonumber\\
&&\quad+\llbracket{\boldsymbol{\varepsilon}}'{}^*,{\boldsymbol{\varepsilon}}\rrbracket
\left(a_{2,1}\llbracket\mathbf{S}_i,\mathbf{S}_j\rrbracket+a_{2,2}\left\llbracket\mathbf{S}_i\times\mathbf{S}_j,\mathbf{S}_k\right\rrbracket\right.\nonumber\\
&&\left.\qquad\qquad\qquad+a_{2,3}\left\llbracket\mathbf{S}_i,\mathbf{S}_j\right\rrbracket\times\mathbf{S}_k\right).\nonumber
\end{eqnarray}


\section{Lower symmetry}
\label{sec:lower-symmetry}

It has been pointed out \cite{haverkort2010b} that, even when
spin-orbit coupling is negligible in the low-energy manifold,
spin-orbit is always very strong in core levels, and may lead to
anisotropies in the RIXS signal. The derivation provided in the main
text is readily generalized to the case of discrete ``spin''
symmetries. 

With the help of the tables found in
Refs.~\onlinecite{bell1954,koster1963,snoke,goss}, one may build bases
for the representations, generalizing those of rotationally-invariant
systems. The formula which generalizes Eq.~\eqref{eq:13} is:
\begin{eqnarray}
  \label{eq:21}
  T_{\mathbf{R}}&=&\sum_\Gamma\sum_l\alpha_{\Gamma,l}\mathcal{E}^{\Gamma,l}\cdot\mathcal{S}^{\Gamma,l},
\end{eqnarray}
where the sum proceeds over all irreducible representations $\Gamma$ of the
point symmetry group at site $\mathbf{R}$ (that of the core hole), $l$
indexes the multiplicity of the representation $\Gamma$, and the dot product represents a symmetric contraction of all
indices. 

\section{Higher multipoles}
\label{sec:higher-multipoles}

As mentioned in the main text, the Hamiltonian at first order in
$\mathbf{A}$ is actually
\begin{equation}
  \label{eq:7}
  H'=\sum_\mathbf{r}\left[\hat{\psi}_\mathbf{r}^\dagger \,\frac{e\mathbf{p}}{m}\,\hat{\psi}_\mathbf{r}\cdot\mathbf{A}+\hat{\psi}_\mathbf{r}^\dagger\,\frac{e\hbar{\boldsymbol{\sigma}}}{2m}\,\hat{\psi}_\mathbf{r}\cdot\left({\boldsymbol{\nabla}}\times\mathbf{A}_\mathbf{r}\right)\right].
\end{equation}
In the main text, only the first term was considered. In the spirit of the derivation provided in the main
text, where experimental parameters are kept explicit, to treat the
second term, one should
consider the couplings to $\mathbf{k}\times{\boldsymbol{\varepsilon}}$
and $\mathbf{k}'\times{\boldsymbol{\varepsilon}}'$, much as we did to
${\boldsymbol{\varepsilon}}$ and ${\boldsymbol{\varepsilon}}'$ upon
considering the term linear in $\mathbf{p}$. One should also note that
``higher multipoles'' will also arise from the expansion of the
exponential, $e^{i\mathbf{k}\cdot\delta\mathbf{r}}=1+i\mathbf{k}\cdot\delta\mathbf{r}-\frac{1}{2}(\mathbf{k}\cdot\delta\mathbf{r})^2+\cdots$.

\section{Details of the cross-section derivations}
\label{sec:deta-cross-sect}

\subsection{Spin nematic in $S=1$ triangular magnets}
\label{sec:spin-nematic}

Following Ref.~\onlinecite{tsunetsugu2006} (the calculation is
performed there in the antiferroquadrupolar phase), we introduce the bosonic
operators $\alpha_\mathbf{r}$ and $\beta_\mathbf{r}$, and the Fock
space vacuum such that
\begin{equation}
  \label{eq:26}
  \begin{cases}
|S^z_\mathbf{r}=0\rangle=|{\rm vac}\rangle\\
|S^z_\mathbf{r}=\pm1\rangle=\frac{1}{\sqrt{2}}(\alpha^\dagger_\mathbf{r}\pm
i\beta^\dagger_\mathbf{r})|{\rm vac}\rangle
\end{cases},
\end{equation}
and
\begin{equation}
  \label{eq:32}
  \begin{cases}
S^x_\mathbf{r}=\alpha^\dagger_\mathbf{r}+\alpha_\mathbf{r}\\
S^y_\mathbf{r}=\beta^\dagger_\mathbf{r}+\beta_\mathbf{r}\\
S^z_\mathbf{r}=-i(\alpha_\mathbf{r}^\dagger\beta_\mathbf{r}-\beta_\mathbf{r}^\dagger\alpha_\mathbf{r})
\end{cases},
\end{equation}
with the constraint that there should be no more than one boson per
site. This in particular implies, in real space:
\begin{equation}
  \label{eq:33}
  \begin{cases}
  \alpha_\mathbf{r}^2=\beta_\mathbf{r}^2=\alpha_\mathbf{r}\beta_\mathbf{r}=\beta_\mathbf{r}\alpha_\mathbf{r}=0\\
\alpha_\mathbf{r}\beta_\mathbf{r}^\dagger=\beta_\mathbf{r}\alpha_\mathbf{r}^\dagger=0\\
\alpha_\mathbf{r}\alpha_\mathbf{r}^\dagger=\beta_\mathbf{r}\beta_\mathbf{r}^\dagger=1-\alpha_\mathbf{r}^\dagger\alpha_\mathbf{r}-\beta_\mathbf{r}^\dagger\beta_\mathbf{r}
\end{cases}.
\end{equation}
Furthermore,
\begin{equation}
  \label{eq:34}
  (\mathbf{S}_i\cdot\mathbf{S}_j)^2=-\frac{1}{2}\mathbf{S}_i\cdot\mathbf{S}_j+\frac{1}{4}\sum_{\mu,\nu}\{S_i^\mu,S_i^\nu\}\{S_j^\mu,S_j^\nu\},
\end{equation}
and
\begin{eqnarray}
  \label{eq:35}
&&\frac{1}{4}\sum_{\mu,\nu}\{S_i^\mu,S_i^\nu\}\{S_j^\mu,S_j^\nu\}\\
&&\qquad\qquad=\sum_\mu(S_i^\mu)^2(S_j^\mu)^2+\frac{1}{2}\sum_{\nu>\mu}\{S_i^\mu,S_i^\nu\}\{S_j^\mu,S_j^\mu\},\nonumber
\end{eqnarray}
with
\begin{equation}
  \label{eq:36}
   \begin{cases}
\{S_\mathbf{r}^x,S_\mathbf{r}^y\}=\alpha_\mathbf{r}^\dagger\beta_\mathbf{r}+\beta_\mathbf{r}^\dagger\alpha_\mathbf{r}\\
\{S_\mathbf{r}^x,S_\mathbf{r}^z\}=-i(\beta_\mathbf{r}-\beta_\mathbf{r}^\dagger)\\
\{S_\mathbf{r}^y,S_\mathbf{r}^z\}=-i(-\alpha_\mathbf{r}+\alpha_\mathbf{r}^\dagger)\\
(S_\mathbf{r}^x)^2=1-\beta_\mathbf{r}^\dagger\beta_\mathbf{r}\\
(S_\mathbf{r}^y)^2=1-\alpha_\mathbf{r}^\dagger\alpha_\mathbf{r}\\
(S_\mathbf{r}^z)^2=\alpha_\mathbf{r}^\dagger\alpha_\mathbf{r}+\beta_\mathbf{r}^\dagger\beta_\mathbf{r}
  \end{cases}
\end{equation}
Using the rules Eq.~\eqref{eq:33} and then keeping only terms quadratic in the boson operators
$\alpha_\mathbf{r}$, $\alpha^\dagger_\mathbf{r}$, $\beta_\mathbf{r}$
and $\beta^\dagger_\mathbf{r}$ (i.e.\ neglecting interactions between
the bosons), we arrive at
\begin{eqnarray}
  \label{eq:37}
    H&=&\frac{1}{2}\left(J_1-J_2\right)\sum_{\eta=\alpha,\beta}\sum_\mathbf{r}\sum_n\left[\eta^\dagger_\mathbf{r}\eta^\dagger_{\mathbf{r}+\mathbf{R}_n}+\eta_\mathbf{r}\eta_{\mathbf{r}+\mathbf{R}_n}\right]\nonumber\\
&&+\frac{J_1}{2}\sum_{\eta=\alpha,\beta}\sum_\mathbf{r}\sum_n\left[\eta^\dagger_\mathbf{r}\eta_{\mathbf{r}+\mathbf{R}_n}+\eta_\mathbf{r}\eta^\dagger_{\mathbf{r}+\mathbf{R}_n}\right]\nonumber\\
&&-\frac{J_2}{2}\sum_{\eta=\alpha,\beta}\sum_\mathbf{r}\sum_n
\left[\eta^\dagger_\mathbf{r}\eta_\mathbf{r}+\eta^\dagger_{\mathbf{r}+\mathbf{R}_n}\eta_{\mathbf{r}+\mathbf{R}_n}\right]\nonumber\\
&=&\frac{\gamma_\mathbf{k}}{2}\left(J_1-J_2\right)\sum_{\eta=\alpha,\beta}\sum_\mathbf{k}\left[\eta^\dagger_\mathbf{k}\eta^\dagger_{-\mathbf{k}}+\eta_\mathbf{k}\eta_{-\mathbf{k}}\right]\nonumber\\
&&+(J_1 \gamma_\mathbf{k}-6J_2)\sum_{\eta=\alpha,\beta}\sum_\mathbf{k}\eta^\dagger_\mathbf{k}\eta_{\mathbf{k}},
\end{eqnarray}
where $\gamma_\mathbf{k}=2(\cos
k_x+\cos(\frac{1}{2}(k_x+\sqrt{3}k_y))+\cos(\frac{1}{2}(k_x-\sqrt{3}k_y)))$. With
the Bogoliubov transformation
$\eta_\mathbf{k}^\dagger=\cosh\xi_\mathbf{k}\rho^\dagger_\mathbf{k}+\sinh\xi_\mathbf{k}\rho_{-\mathbf{k}}$,
we obtain
\begin{equation}
  \label{eq:38}
  H=\sum_{\rho=\rho^\alpha,\rho^\beta}\sum_\mathbf{k}\omega_\mathbf{k}\rho^\dagger_\mathbf{k}\rho_\mathbf{k},
\end{equation}
where we have defined:
\begin{equation}
  \label{eq:42}
\omega_\mathbf{k}=2[A_\mathbf{k}\cosh2\xi_\mathbf{k}+B_\mathbf{k} \sinh2\xi_\mathbf{k}]=2\sqrt{A_\mathbf{k}^2-B_\mathbf{k}^2},
\end{equation}
with
\begin{equation}
  \label{eq:39}
  \begin{cases}
A_\mathbf{k}=\frac{1}{2}(J_1\gamma_\mathbf{k}-6J_2)\\
B_\mathbf{k}=\frac{\gamma_\mathbf{k}}{2}(J_2-J_1)
\end{cases}
\end{equation}
if
\begin{equation}
  \label{eq:41}
  \begin{cases}
A_\mathbf{k}\sinh2\xi_\mathbf{k}+B_\mathbf{k}\cosh2\xi_\mathbf{k}=0\\
(A_\mathbf{k}\cosh2\xi_\mathbf{k}+B_\mathbf{k}\sinh2\xi_\mathbf{k})^2=A_\mathbf{k}^2-B_\mathbf{k}^2
\end{cases}.
\end{equation}
This yields:
\begin{equation}
  \label{eq:40}
  \begin{cases}
\sinh^22\xi_\mathbf{k}=\frac{B_\mathbf{k}^2}{A_\mathbf{k}^2-B_\mathbf{k}^2}\\
\cosh^22\xi_\mathbf{k}=\frac{A_\mathbf{k}^2}{A_\mathbf{k}^2-B_\mathbf{k}^2}
\end{cases}.
\end{equation}
Since $\forall x$ $\cosh x>0$, 
\begin{equation}
  \label{eq:43}
  \cosh2\xi_\mathbf{k}=\sqrt{\frac{A_\mathbf{k}^2}{A_\mathbf{k}^2-B_\mathbf{k}^2}},\quad
\sinh2\xi_\mathbf{k}=-\frac{B_\mathbf{k}}{A_\mathbf{k}}\sqrt{\frac{A_\mathbf{k}^2}{A_\mathbf{k}^2-B_\mathbf{k}^2}}.
\end{equation}

The transition operator Eq.~\eqref{eq:15} takes the form, in Fourier
space:
\begin{eqnarray}
  \label{eq:44}
  T_\mathbf{k}&=&\kappa^{(0)}\delta(\mathbf{k})-\sum_\mu\sum_{\rho=\rho^\mu}\left[\rho^\dagger_{\mathbf{k}}(\mathcal{A}_\mu\cosh\xi_\mathbf{k}+\mathcal{B}_\mu\sinh\xi_\mathbf{k})\right.\nonumber\\
&&\qquad\qquad+\rho_{-\mathbf{k}}(\mathcal{A}_\mu\sinh\xi_\mathbf{k}+\mathcal{B}_\mu\cosh\xi_\mathbf{k})\Big],
\end{eqnarray}
where
\begin{eqnarray}
  \label{eq:45}
  \kappa^{(0)}&=&\alpha_0({\boldsymbol{\varepsilon}}\cdot{\boldsymbol{\varepsilon}}'{}^*)\\
\kappa_\mu^{(1)}&=&\alpha_1\epsilon_{\mu\sigma\tau}\varepsilon^\sigma\varepsilon'{}^*{}^\tau\\
\kappa_{\mu\nu}^{(2)}&=&\alpha_2\left[-\frac{2}{3}\delta_{\mu\nu}({\boldsymbol{\varepsilon}\cdot{\boldsymbol{\varepsilon}}'{}^*})+\varepsilon^\mu\varepsilon'{}^*{}^\nu+\varepsilon^\nu\varepsilon'{}^*{}^\mu\right]\\
\mathcal{A}_x&=&\kappa_{xy}^{(2)}+i\kappa_z\\
\mathcal{A}_x&=&\kappa_{yz}^{(2)}-i\kappa_x\\
\mathcal{B}_x&=&\kappa_{xy}^{(2)}-i\kappa_z\\
\mathcal{B}_x&=&\kappa_{yz}^{(2)}+i\kappa_x.
\end{eqnarray}
Plugging this into the expression for the cross section:
\begin{eqnarray}
  \label{eq:46}
  \frac{\delta^2\sigma}{\delta\Omega\delta\omega}&\propto&\sum_{\mu=x,z}\left|\mathcal{A}_\mu\cosh\xi_\mathbf{k}+\mathcal{B}_\mu\sinh\xi_\mathbf{k}\right|^2\delta(\omega-\omega_\mathbf{k}),\nonumber
\end{eqnarray}
we arrive at the result given in the main text.

\subsection{Vector chirality and bond nematic in $S=1/2$ $J_1-J_2$ chains}
\label{sec:vect-chir-bond}

Equal-time and real-space correlation functions are given in
Refs.~\cite{mcculloch2008,hikihara2008}. Here, we find the following
contributions to the cross section:
\begin{itemize}
\item in the nematic phase:
  \begin{eqnarray}
    \label{eq:47}
&&    \mathcal{I}^{\langle
   (S^+S^+)_{-\omega,-k}(S^+S^+)_{\omega,k}\rangle}\\
&&\quad\propto\mathcal{A}\sum_{\epsilon=\pm1}\frac{\Theta(\omega^2-v_+^2(k-\epsilon\pi)^2)}{\sqrt{\omega^2-v_+^2(k-\epsilon\pi)^2}^{2-1/K_+}}\nonumber\\
&&\qquad+\mathcal{B}\sum_{\epsilon,\epsilon'}\Theta\left[\omega^2-v_+^2\left(k-\pi(\epsilon\frac{1}{2}-\epsilon'M)\right)^2\right]\nonumber\\
&&\qquad\qquad\times\sqrt{\omega^2-v_+^2\left(k-\pi(\epsilon\frac{1}{2}-\epsilon'M)\right)^2}^{K_++1/K_+-2}\nonumber
\end{eqnarray}
\begin{eqnarray}
  \label{eq:48}
  \mathcal{I}^{\langle S^+_{-\omega,-k}S^-_{\omega,k}\rangle}&=&{\rm gapped}\\
  \label{eq:49}
  \mathcal{I}^{\langle\chi^{z}_{-\omega,-k}\chi^z _{\omega,k}\rangle}&\propto&\omega\left(\delta(\omega+v_+k)+\delta(\omega-v_+k)\right)
\end{eqnarray}

\begin{eqnarray}
  \label{eq:50}
  \mathcal{I}^{\langle S^z_{-\omega,-k}S^z_{\omega,k}\rangle}
&\propto&\omega\left(\delta(\omega+v_+k)+\delta(\omega-v_+k)\right)\\
&&+\sum_{\epsilon=\pm1}\frac{\Theta(\omega^2-v_+^2(k-\epsilon\pi(\frac{1}{2}-M))^2)}{\sqrt{\omega^2-v_+^2(k-\epsilon\pi(\frac{1}{2}-M))^2}^{2-K_+}}\nonumber
\end{eqnarray}

\item in the vector chiral phase:
  \begin{eqnarray}
    \label{eq:51}
&&\mathcal{I}^{\langle\chi^{z}_{-\omega,-k}\chi^z _{\omega,k}\rangle}\\
&&\quad\propto
\mathcal{A}\omega^3\left(\delta(\omega+v_+k)+\delta(\omega-v_+k)\right)\nonumber\\
&&\qquad+\mathcal{B}\sum_{\epsilon,\epsilon'=\pm1}\left[\Theta(\omega^2-v_+^2(k-\epsilon2\pi
M-\epsilon'\pi)^2)\right]\nonumber\\
&&\qquad\qquad\qquad\times\sqrt{\omega^2-v_+^2(k-\epsilon2\pi
          M-\epsilon'\pi)^2}^{4K_+-2}\nonumber
\end{eqnarray}
\begin{eqnarray}
  \label{eq:52}
 \mathcal{I}^{\langle S^z_{-\omega,-k}S^z_{\omega,k}\rangle}&\propto&
\omega\left(\delta(\omega+v_+k)+\delta(\omega-v_+k)\right)\\
\mathcal{I}^{\langle S^x_{-\omega,-k}S^x_{\omega,k}\rangle}
&\propto&\sum_{\epsilon=\pm1}\frac{\Theta(\omega^2-v_+^2(k-\epsilon
              Q)^2)}{\sqrt{\omega^2-v_+^2(k-\epsilon
              Q)^2}^{2-1/(4K_+)}}\nonumber
\end{eqnarray}
  \begin{eqnarray}
    \label{eq:53}
   &&\mathcal{I}^{\langle
   (S^+S^+)_{-\omega,-k}(S^+S^+)_{\omega,k}\rangle}\\
&&\qquad\qquad\propto\sum_{\epsilon=\pm1}\frac{\Theta(\omega^2-v_+^2(k-\epsilon2Q)^2)}{\sqrt{\omega^2-v_+^2(k-\epsilon2Q)^2}^{2-1/K_+}}.\nonumber
  \end{eqnarray}
\end{itemize}
In all the above, $\mathcal{A}$ and $\mathcal{B}$ are constants, and $Q=\frac{\pi}{2}-\frac{1}{2}\sqrt{\frac{\pi}{2}}\langle\partial_x\theta_+\rangle$.

Note, in particular, that, since $K$ increases monotonically between
$1/2$ and $1$ ($K(M=0)=1/2$ and $K(M=1/2)=1$), and
$K_+=K(1+K\frac{J_1}{\pi v})$, $K_+\geq 1/2$. Moreover, the
bosonization approach is valid only ``not too close'' from the
saturation limit $M=1/2$, and in the weak coupling regime $v\sim
J_2$. So, in particular:
\begin{equation}
  \label{eq:54}
  \begin{cases}
2-\frac{1}{K_+}\geq3/2\\
K_++\frac{1}{K_+}-2\geq0\\
2-K_+\geq0&\mbox{if }K\leq\pi v\frac{-1+\sqrt{1+8J_1/(\pi v)}}{2J_1}\\
4K_+-2\geq0\\
2-\frac{1}{4K_+}\geq\frac{15}{8}
\end{cases}.
\end{equation}
Note that $K\leq\pi v\frac{-1+\sqrt{1+8J_1/(\pi v)}}{2J_1}$ is always
true for $\frac{J_1}{\pi v}\leq1$.

\end{document}